\documentclass[9pt,twocolumn,twoside]{pnas-new}

\templatetype{pnasresearcharticle} 

\usepackage{xspace}

\newcommand{\mj}{\ensuremath{M_\mathrm{J}}\xspace}

\newcommand{\Msun}{\ensuremath{M_{\odot}}\xspace}

\def\deg{\ensuremath{^{\circ}}}

\usepackage{amsmath}
\usepackage{bm}
\usepackage{mathtools}
\usepackage{longtable}
\usepackage{lscape}
\usepackage[flushleft]{threeparttable}
\usepackage{multirow}

\begin{document}

\title{A universal brown dwarf desert formed between planets and stars}

\author[a,b,c]{Kaiming Cui}
\author[a]{Guang-Yao Xiao}
\author[a,d,1]{Fabo Feng}
\author[e,f]{Beibei Liu}
\author[g]{Sergei Nayakshin}
\author[h,i]{Cassandra Hall}
\author[a]{Kangrou Guo}
\author[a,j]{Dong Lai}
\author[a,d]{Masahiro Ogihara}
\author[a]{Yicheng Rui}
\author[k]{Alan P. Boss}
\author[k]{R. Paul Butler}
\author[a]{Yifan Xuan}

\affil[a]{Tsung-Dao Lee Institute, Shanghai Jiao Tong University, 1 Lisuo Road, Shanghai 201210, People's Republic of China}
\affil[b]{Department of Physics, University of Warwick, Gibbet Hill Road, Coventry CV4 7AL, UK}
\affil[c]{Centre for Exoplanets and Habitability, University of Warwick, Gibbet Hill Road, Coventry CV4 7AL, UK}
\affil[d]{School of Physics and Astronomy, Shanghai Jiao Tong University, 800 Dongchuan Road, Shanghai 200240, People's Republic of China}
\affil[e]{Institute for Astronomy, School of Physics, Zhejiang University, Hangzhou 310027, People's Republic of China}
\affil[f]{Center for Cosmology and Computational Astrophysics, Institute for Advanced Study in Physics, Zhejiang University, Hangzhou 310027, People's Republic of China}
\affil[g]{School of Physics and Astronomy, University of Leicester, Leicester LE1 7RH, UK}
\affil[h]{Department of Physics and Astronomy, The University of Georgia, Athens, GA 30602, USA}
\affil[i]{Center for Simulational Physics, The University of Georgia, Athens, GA 30602, USA}
\affil[j]{Department of Astronomy, Cornell Center for Astrophysics and Planetary Science, Cornell University, Ithaca, NY 14853, USA}
\affil[k]{Earth and Planets Laboratory, Carnegie Institution for Science, 5241 Broad Branch Road, NW, Washington, DC 20015, USA}
\leadauthor{Cui}


\significancestatement{Giant planets and brown dwarfs represent crucial transition objects between planets and stars, yet their formation mechanisms remain uncertain. Using a sample of 55 companions detected through combined radial velocity and astrometry, we identify a universal brown dwarf desert at approximately 30 Jupiter masses extending to at least 20 astronomical units. This desert separates two distinct populations formed through different mechanisms: lower-mass companions likely formed by core accretion, while higher-mass companions formed by gravitational instability. {Our findings provide observational evidence that the existence of the brown dwarf desert results from these two distinct formation pathways.}}

\authorcontributions{F.F. designed research; K.C.,
G.-Y.X., B.L., S.N., K.G., Y.R., A.P.B., R.P.B., and Y.X.
performed research; K.C. and G.-Y.X. analyzed data; and
K.C., G.-Y.X., F.F., B.L., S.N., C.H., K.G., D.L., M.O., and Y.R.
wrote the paper.}

\authordeclaration{The authors declare no conflict of interest.}
\correspondingauthor{\textsuperscript{1}To whom correspondence should be addressed. E-mail: ffeng@sjtu.edu.cn}


\begin{abstract}
Giant planets and brown dwarfs play a crucial role in star and planet formation 
, as they are situated at the boundary between planets and stars with uncertain formation mechanisms. Previous observational searches for the formation boundary 
were hampered by the lack of large unified samples of wide-orbit giant planets and substellar companions. 
A combined analysis of radial velocity and astrometry mitigates this problem and has significantly enlarged the sample. Here we present a rigorous statistical analysis of the sample of 55 giant planets, brown dwarfs and low-mass stellar companions orbiting FGK stars. 
We quantitatively analyze the occurrence rates of brown dwarfs and identify a distinct brown dwarf desert at approximately $30\,\mj$, with no evidence of disappearance up to 20\,au. {Unlike previous studies that predicted a declining planet occurrence rate beyond the water-ice line 
, we identify a new population of giant planets and low-mass brown dwarfs in this region.} The metallicity and eccentricity trends in our sample suggest that these are the consequences of two different formation scenarios. Our combined population synthesis model successfully accounts for the observed brown dwarf desert, supporting the dual formation hypothesis.
\end{abstract}

\doi{\url{https://doi.org/10.1073/pnas.2524764123}}

\maketitle
 \thispagestyle{firststyle}
\ifthenelse{\boolean{shortarticle}}{\ifthenelse{\boolean{singlecolumn}}{\abscontentformatted}{\abscontent}}{}



The formation and evolution of planetary systems is a central issue in astrophysics, with giant planets and brown dwarfs occupying a particularly intriguing niche in this context. These substellar objects straddle the boundary between the most massive planets and the least massive stars, making them key laboratories for testing theories of both star and planet formation. The mechanisms behind their formation remain largely uncertain, as traditional paradigms like core accretion for planets and gravitational collapse for stars may not apply uniformly across all giant planets and brown dwarfs. Understanding where and how exactly the transition occurs between planetary-mass companions and brown dwarfs, and the processes responsible for their birth, is crucial for assembling a comprehensive theory of how planetary systems arise and evolve \cite{idaDeterministicModelPlanetary2004, chabrierGiantPlanetBrown2014, helledGiantPlanetFormation2014, dangeloFormationGiantPlanets2018, drazkowskaPlanetFormationTheory2023}.

A particularly puzzling feature in this context is the so-called “brown dwarf desert,” a well-studied dearth of brown dwarf companions to Sun-like stars at close orbital separations \cite{campbell88,gretherHowDryBrown2006,maStatisticalPropertiesBrown2014,shahafStudyMassratioDistribution2019,ungerExploringBrownDwarf2023,chenProbingShapeBrown2024}. Despite their pivotal role in planetary system architecture, efforts to explore the giant planet-brown dwarf formation boundary, including the brown dwarf desert region, have faced significant challenges. Observational searches to precisely depict this boundary, including the census of wide-orbit giant planets and substellar companions, have historically been limited by the sparsity and incompleteness of large and homogeneous samples. Efforts using techniques like direct imaging and radial velocity surveys have provided valuable insights \cite{fengDetectionNearestJupiter2019, nielsenGeminiPlanetImager2019, bowlerPopulationlevelEccentricityDistributions2020, viganSPHEREInfraredSurvey2021, fultonCaliforniaLegacySurvey2021}, but progress has been slow due to instrumental limitations and the relative rarity of such companions at wide separations.

In recent years, advances in the field have been driven by the advent of high-precision astrometric observations. The European Space Agency’s Gaia mission, through its successively more accurate and comprehensive data releases \cite{gaia16, gaia22, Gaia2023DR3}, has revolutionized our ability to detect the tiny positional wobbles of stars induced by orbiting substellar companions. When Gaia data are combined with the historical baseline provided by the Hipparcos mission \cite{perryman97, leeuwen07}, the resulting multi-decade astrometric timelines {will} dramatically enhance the detectability of giant planets and brown dwarfs, especially in the regime of wide orbits that are challenging for other techniques.

Crucially, the integration of high-precision astrometry with complementary methods such as radial velocity monitoring has opened new frontiers. Through joint analyses that combine these approaches, astronomers have achieved more robust detections, improved constraints on orbits and masses, and a better ability to distinguish between planetary and substellar companions in parameter space \cite{brandt19, kervella22, feng3DSelection1672022,gilbertOrbitalEccentricitiesSuggest2025,zandtSmoothTransitionGiant2025}. This synergy represents a major leap forward for constructing statistically significant samples of wide-orbit companions, enabling the first systematic explorations of the factors shaping the formation and evolution of giant planets and brown dwarfs.

As a result, the current era marks an unprecedented opportunity to address long-standing questions about the nature of the planet-brown dwarf divide and the diversity of planetary system architectures. In this paper, we present a demographics study of planetary and substellar companions ($5\text{--}120\mj$) on wide orbits (2--20\,au) around late FGK-type main-sequence stars, leveraging the latest high-precision data to advance our understanding of brown dwarf desert.

Here, we analyze a sample of 55 planetary and substellar companions ($5\text{--}120\mj$) on wide orbits (2--20\,au) around late FGK-type main-sequence stars, selected from a sample of companions detected through combined radial velocity and astrometric analyses \cite{feng3DSelection1672022}. The selection considers the observation bias, stellar types, reliability of stellar parameters, and detection algorithms.

\section*{Methods}\label{sec:methods}
\begin{figure}
\centering
\includegraphics[width=\columnwidth]{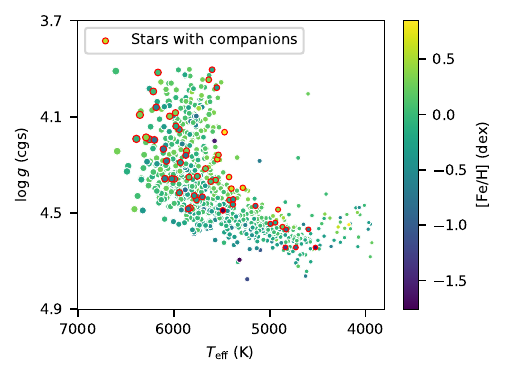}
\caption{The Hertzsprung-Russell (HR) diagram of our final selected stars color-coded by the stellar metallicity. The circle size is proportional to the stellar mass. }
\label{fig:stellar sample}
\end{figure}

We begin with a high-level summary of our methods before providing more details in the sections below. We first select a sample of 790 FGK stars hosting 55 companions (see Fig.~\ref{fig:stellar sample}), drawn from our previous detection results. The selection was designed to construct a less biased sample for demographic studies. We quantify our detection efficiency and reliability through injection-and-recovery tests. We then estimate the occurrence rates as a function of companion mass and orbital semi-major axis. Finally, we perform population synthesis modeling to explain the observed occurrence rates.

\subsection*{Sample selection}
The sample used in this study is drawn from a catalog of 167 substellar companions, with masses ranging from 5 to 120 $\mj$, detected by ref. \cite{feng3DSelection1672022}. These detections are based on a combined analysis of radial velocity and astrometric data from Hipparcos and Gaia, covering 5,108 AFGKM stars. The catalog was generated through a blind search of all available data for long-period radial velocity signals, which are effectively constrained by the combined Gaia and Hipparcos observations.

However, it is not appropriate to use this catalog directly, as its observations come from multiple sources, making the sample heterogeneous. Ideally, a complete survey would require a volume-limited or magnitude-limited target selection; however, this is typically unfeasible for RV surveys due to observational strategies and associated costs.

As a compromise, it is common practice to construct a sample that combines several different surveys known to be less biased \cite{rosenthalCaliforniaLegacySurvey2021}. Our sample consists of targets from four long-term RV surveys: the HARPS Guaranteed Time Observations (GTO) programs \cite{mayor03}, the California Planet Search \citep[CPS;][]{Howard2010ApJ...721.1467H}, the Anglo-Australian Planet Search \citep[AAPS;][]{Jones2002MNRAS.337.1170J, Tinney2003ApJ...587..423T}, and the Magellan Planet Search Program \cite{Arriagada2011ApJ...734...70A}. These surveys employ a more uniform selection process compared to dedicated follow-up campaigns for transiting planets or other special-interest targets. The detailed characteristics of those four surveys are available in {Supplemental Information}.

We first cross-matched our original sample of 5,108 stars with these four surveys, this means selecting objects in those four surveys with at least five RV measurements. After applying the stellar mass cut ($0.6<M_*<1.4\,\Msun$) and $\log g > 3.9$, we ultimately selected {790} stars, among which {55} substellar companions were identified orbiting {54} stars. They are listed in the Supplemental Information.
The stellar properties, such as {stellar }mass ($M_{\star}$), effective temperature ($T_{\rm eff}$), surface gravity (log\,$g$), and metallicity ([Fe/H]), are adopted from the TESS Input Catalog (TIC) version 8.2 \cite{Paegert2021}, except for the metallicity of the {54} companion-hosting stars. We determine their metallicities based on spectroscopic surveys, with further details in the \hyperlink{si}{Supplemental Information}.

\subsection*{Injection of signals}

To assess detection efficiency for each star using RV data, we adopt the injection-recovery technique. For systems with previously identified planetary companions, we subtract the best-fit Keplerian model from the observed RV time series to obtain residuals that primarily represent stellar or instrumental noise. For all systems, including those without detected companions, we inject synthetic planet-induced RV signals into these residuals, simulating a wide range of planetary masses and orbital parameters. The synthetic RV signal at each observation epoch is calculated using the Keplerian formula:
\[
{\rm RV} = K\,[{\rm cos}(\omega+\nu)+e\,{\rm cos}(\omega)],
\]
where $K$ is the velocity semi-amplitude, $\omega$ is the argument of periastron, $\nu$ is the true anomaly at time $t$, and $e$ is the eccentricity. Orbital parameters are sampled uniformly over broad ranges, except for those fixed by each system {(i.e., the time of periastron passage $T_p$ and stellar mass $M_\star$)}, enabling a comprehensive exploration of detection thresholds. {We inject only one companion at a time, so our injections do not include multi-companion systems.}

The astrometric injection follows a similar framework. 
We generate synthetic astrometric signatures by adding simulated planet-induced stellar reflex motions to the astrometric time series. The synthetic signal is computed for each epoch using the projected two-dimensional {(2D)} orbit on the sky, accounting for the system's orientation and orbital parameters (such as mass ratio, orbital period, inclination, ascending node, and argument of periastron). 
{Then we project the 2D stellar offsets onto the 1D along-scan direction based on the scan law of Hipparcos and Gaia satellites.}
The astrometric displacement at each epoch is then added to the residual abscissa, and the data are analyzed to determine whether the injected signal can be recovered with the available astrometric precision and cadence. 

\subsection*{Recovery of signals}
Given the injected signals in both the RV and astrometry data, we recover these signals using the same detection pipeline applied to the observed sample. The key step is to identify RV signals through Keplerian model fitting, requiring $\ln \text{BF} > 5$, along with additional criteria for orbital periods longer than the observational baseline. Once the RV signal is deemed significant, we perform joint fitting of the synthesized Hipparcos and Gaia astrometry. {In order to match the real discoveries, }our final set of recovered companions is limited to those with relative planet mass errors less than {60\%}. Please see the \hyperlink{si}{Supplemental Information} for details.

\subsection*{Detection efficiency and reliability}
For each star, we inject {75,000} trial signals and repeat above recovery procedure. We make a grid of bins in our 4-dimension parameter space ($a\text{--}P\text{--}m\text{--}q$), and count the numbers of recovers divided by the injected sources as the detection efficiency. The individual detection efficiency can be combined to illustrate the overall detection capabilities of the survey. {Because each of our injected systems hosts only one companion, this may introduce a potential bias when compared with our observations. However, since the companions we detect usually have much longer periods than inner companions, this bias is likely to be insignificant.}

{Unlike the efficiency analysis that applies to all 790 stars, we restrict the reliability study to the 54 companion-hosting systems.}
The reliability is calculated by dividing the number of valid signals by the total recovered signals in each bin. We additionally require that the valid signals should satisfy: 1) the relative differences for semi-amplitude $K$, period $P$ {and companion mass $m$} between injected and recovered signal are less than 30\%, 2) the eccentricity differences are within 0.1, and 3) the inclination differences are within $30\deg$. Almost all the quantities across various parameters are larger than 50\%, suggesting the advantages and reliability of our detection method. Our detection efficiency and reliability are illustrated in \hyperlink{si}{Supplemental Information}.

\subsection*{Occurrence rate estimation}
We first adopt a detection efficiency-weighted KDE (wKDE) to qualitatively demonstrate the 2D occurrence distribution without binning effect. We then apply the hierarchical Bayesian methodology \cite{hoggINFERRINGECCENTRICITYDISTRIBUTION2010, youdinExoplanetCensusGeneral2011,foreman-mackeyEXOPLANETPOPULATIONINFERENCE2014} to model the occurrence rate for both 1D and 2D parameter spaces.

The occurrence rate density is defined as 
\begin{equation}
    \Gamma = \frac{\mathrm{d}^2N}{\mathrm{d}\ln{m}\ \mathrm{d}\ln{a}} ,
\end{equation}
where $m$ and $a$ are companion mass in the unit of $\mj$ and semi-major axis in the unit of au. {The occurrence rate is obtained by integrating the occurrence rate density over the relevant parameter space.}
Considering the detection efficiency $ Q(w)$, the observable occurrence rate {density} $\hat{\Gamma}$ is 
\begin{equation}
   \hat{\Gamma}_{\boldsymbol{\theta}} (\boldsymbol{w}) = Q(\boldsymbol{w}) \Gamma_{\boldsymbol{\theta}} (\boldsymbol{w}),
\end{equation}
where the $\boldsymbol{\theta}$ are the parameters used to describe the occurrence rate model. $\boldsymbol{w}$ are the physical parameters (semi-major axis $a$ and companion mass $m$ in this work).

Taking into account observational uncertainties, the likelihood can be written as 
\begin{equation}
    \mathcal{L} \approx \exp\left( -\int{\hat{\Gamma}_{\boldsymbol{\theta}}(\boldsymbol{w})\mathrm{d}\boldsymbol{w}} \right) \prod_{k=1}^K \frac{1}{N_k}\sum_{n=1}^{N_k}\frac{\hat{\Gamma}_{\boldsymbol{\theta}}(\boldsymbol{w}_k^{(n)})}{p(\boldsymbol{w}_k^{(n)})}.
\end{equation}
$N_k$ represents the number of posterior samples corresponding to the $k$-th companion. The $\boldsymbol{w}_k^{(n)}$ denotes the $n$-th sample associated with the same $k$-th companion. Furthermore, $p(\boldsymbol{w}_k^{(n)})$ signifies the prior probability distribution of each parameter's posteriors for every individual companion. They are taken from the our previous MCMC results \cite{feng3DSelection1672022}.

We define the non-parametric occurrence rate as a step function with n-bins
\begin{equation} \Gamma(\boldsymbol{w}|\boldsymbol{\theta}) = \theta_n | \boldsymbol{w} \in \Delta_n.
\end{equation}
Here, $\theta_n$ signifies the occurrence rate of each bin step $\Delta_n$. In different contexts, the symbol $\boldsymbol{w}$ denotes either $a$ or $m$ in 1D scenarios, while encompassing both $a$ and $m$ in 2D scenarios.

\subsection*{Parametric occurrence rate model}
Similar with previous work \cite{foreman-mackeyEXOPLANETPOPULATIONINFERENCE2014,fultonCaliforniaLegacySurvey2021}, parametric models are applied in both 1D and 2D. Here we show our parametric models for different scenarios.

\begin{SCfigure*}
    \centering
    \includegraphics[width=0.6\textwidth]{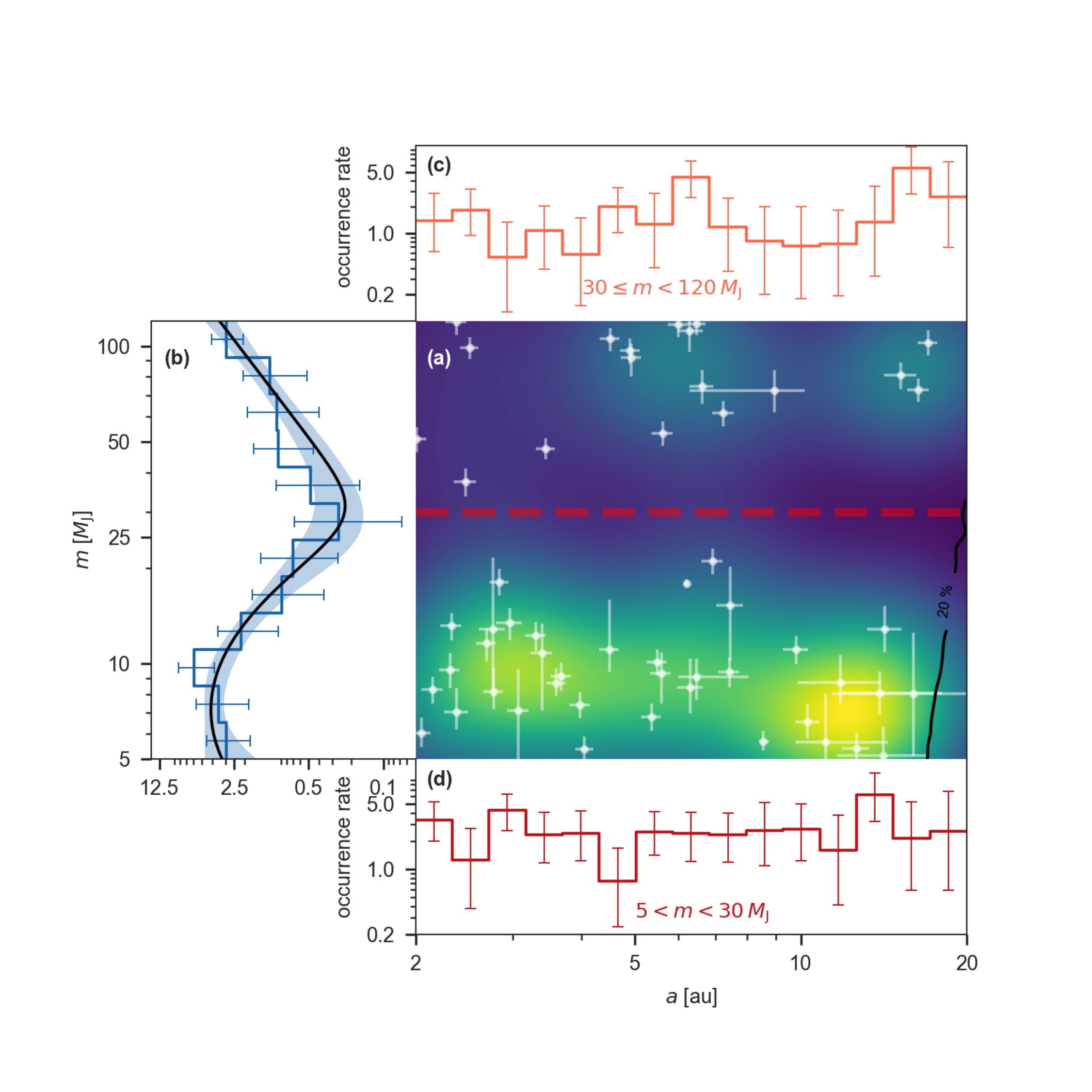}
    \caption{{Distribution of companions in semi-major axis and mass, along with the occurrence rate density of subsamples.} {(a)}, Detection efficiency-weighted KDE and distribution of our sample (white points). We divide the sample into low-mass and high-mass subsamples, with the division marked by a dashed red line at $30\mj$. 
    Detection efficiencies of 20\% is outlined by a black contour. {(b),} The occurrence rate {density} vs companion mass ($\mathrm{d}N/\mathrm{d}\ln m$ per 100 stars). The best-fit log-normal plus power-law model with its uncertainty is overplotted. {The uncertainties correspond to the 16th and 84th percentiles of the posterior distributions.}
    {(c)}, The occurrence rate {density} vs semi-major axes (${\mathrm{d}N}/{\mathrm{d}\ln a}$ per 100 stars), when companion mass above 30\,$\mj$. {(d)}, similar to {c} but for companions with $m < 30$\,$\mj$.
    }
    \label{fig:weighted-KDE}
\end{SCfigure*}

We applied a parametric model to describe the occurrence rate {density} at different masses. The model combines a log-normal distribution with a power law.
\begin{equation}
    \Gamma(m) = \frac{A}{m \sigma}\exp\left(-\frac{(\ln m - \mu)^2}{2\sigma^2}\right) + C m^{\alpha} \label{eq: log-normal-power-law}.
\end{equation}
In this context, $A$ and $C$ are constants, while $\alpha$ represents the power-law indices. The parameters $\mu$ and $\sigma$ characterize a log-normal distribution. The prior distributions for the logarithms of the constants in base 10, $\lg A$ and $\lg C$, are sampled from a uniform distribution denoted as $ \mathcal{U}(-5, 5) $. The priors for $\mu$ and $\sigma$ follow distributions from $ \mathcal{U}(\ln(2), \ln(20)) $ and $ \mathcal{U}(0, 5) $, respectively. Additionally, the prior for the power index is defined as $ \mathcal{U}(0, 5) $. The modeled result is shown in Fig.~\ref{fig:weighted-KDE}b. The posterior corner plot is shown in \hyperlink{si}{Supplemental Information}. 
Another model for comparison is a flat model, the occurrence rate {density} is a constant.
\begin{equation}
    \Gamma(m) = C
\end{equation}
The posterior of the constant is $C=12.4^{+1.1}_{-1.2}$.

Considering that the occurrence models are nested and {may have non-gaussian posteriors, we use the Pareto-smoothed importance sampling leave-one-out cross-validation \citep[PSIS-LOO-CV;][]{JMLR:v25:19-556,vehtariPracticalBayesianModel2017}, which is a very robust model comparison tool. 
The PSIS-LOO estimate of expected log predictive density (ELPD) is calculated as
\begin{equation}
    \text{ELPD LOO} = \sum_{i=1}^n \log \left[ \frac{1}{S} \sum_{s=1}^S w_i^{(s)} p(y_i \mid \theta^{(s)}) \right],
\end{equation}
where $y_i$ is the $i$-th data point, $\theta^{(s)}$ are posterior draws, $S$ is the number of posterior draws, and $w_i^{(s)}$ represent the leave-one-out importance ratios stabilized through Pareto smoothing. The higher ELPD indicates the better model.
The comparison of models can be quantified using the difference of ELPD divided by the standard error of ELPD.}

We also characterize the brown dwarf boundary by following the methodology in ref. \cite{mazehDearthShortperiodNeptunian2016}. Both upper and lower boundaries are lines in the $(\ln a, \ln m)$ plane. The occurrence rate {density} can be modeled as 

\begin{equation}
\begin{aligned}
    \Gamma(d_\mathrm{upper}, d_\mathrm{lower}, \delta_1, \delta_2, \Delta) & = A_1 \left( \frac{1}{1 + \exp{(-d_\mathrm{upper}/\delta_1)}} + \Delta \right) \\
    & + A_2 \left( \frac{1}{1+ \exp{(-d_\mathrm{lower}/\delta_2)}} + \Delta \right)
\end{aligned}
\end{equation}

The variables $d_\mathrm{upper}$ and $d_\mathrm{lower}$ represent the distances from the upper and lower boundaries, respectively. These distances are positive when above the boundary line and negative when below it. {The upper and lower boundaries are represented as lines on a logarithmic scale in $m-a$ plane ($\ln m = \mathbf{k} \ln a + \mathbf{b}$), where $(k_1, b_1)$ and $(k_2, b_2)$ correspond to the upper and lower boundaries, respectively.} The parameters $\delta_1$ and $\delta_2$ denote the transition widths, while $\Delta$ indicates the low density within the desert region. Following the MCMC sampling, the posterior estimates of the boundary locations along with their associated uncertainties are calculated. The posterior distributions of all parameters are shown in \hyperlink{si}{Supplemental Information}.

\subsection*{Combined population synthesis}
Our population synthesis study incorporates two models: core accretion (CA) and gravitational instability (GI). The CA model, despite being the most conventional model for planet formation, primarily yields giant planets but struggles to replicate our observation when used in isolation. Hence, combining CA and GI models can cover a broader region in our parameter space.

For planet population synthesis, we employ both the pebble-driven CA and GI models. The CA model uses a Monte Carlo sampling approach to generate initial disc, planet embryo, and stellar properties for each system, and follows the evolution of embryos via pebble accretion. {We only track the growth and migration of a single protoplanet per system. Therefore, these planets have purely circular and coplanar orbits.} The GI model is based on an analytical treatment of disc fragmentation, with planet migration modeled using standard type I and type II prescriptions. Further details of both models can be found in the \hyperlink{si}{Supplemental Information}.

We integrate the CA model and GI model to replicate our observed distribution of occurrence rates across three dimensions: $a$, $m$, and [Fe/H]. The combined model introduces two additional parameters, $f_1$ and $f_2$, which are designed to regulate the contribution of occurrence rates from both CA and GI models respectively. Thus the combined occurrence rate $\Gamma_\mathrm{comb}$ can be expressed as 
\begin{equation}
    \Gamma_\mathrm{comb} = f_1 D_\mathrm{CA} + f_2 D_\mathrm{GI},
    \label{eq: pop-comb}
\end{equation}
where $D_\mathrm{CA}$ and $D_\mathrm{GI}$ are corresponding number densities of CA and GI simulated samples.
To simplify the occurrence rate model comparison, we bin both the observation and theoretical model samples to $6\times6\times3$ bins in the $a\text{--}m\text{--[Fe/H]}$ space. The binning is uniform in the log scale for $a$ and $m$, and the three bins in metallicity are separated by -0.1 and 0.1 as metal poor ($\text{[Fe/H]}<- 0.1$), solar metallicity ($-0.1\leq \text{[Fe/H]} \leq 0.1$), and metal rich ($\text{[Fe/H]}> 0.1$).
Since the bin size is larger than most $ 1\sigma $ uncertainties of observation, and to simplify the calculation, we estimate their occurrence rates without considering uncertainties. The Poisson process log-likelihood can be expressed as follows:
\begin{equation}
    \ln \mathcal{L} = \sum_{i=1}^{N_b} \left[ k_i \ln \lambda_i(\boldsymbol{\theta}) - \lambda_i(\boldsymbol{\theta}) - \ln(k_i!) \right],
\end{equation}
where $N_b$ represents the total number of bins, $k$ denotes the observed occurrence rate, and $\lambda(\boldsymbol{\theta})$ is the expected rate predicted by population synthesis models with model parameters denoted by $\boldsymbol{\theta}$.

Given the non-linearity of the model, we employ the differential evolution algorithm \cite{stornDifferentialEvolutionSimple1997,bowenCharacterizationStructuresXray1999} for global optimization. Due to the stochastic properties in our PPS results, we performed the optimization ten times iteratively to estimate parameter uncertainties since MCMC is overly heavy for population synthesis simulations. 

\begin{figure}[!h]
    \centering
    \includegraphics[width=\columnwidth]{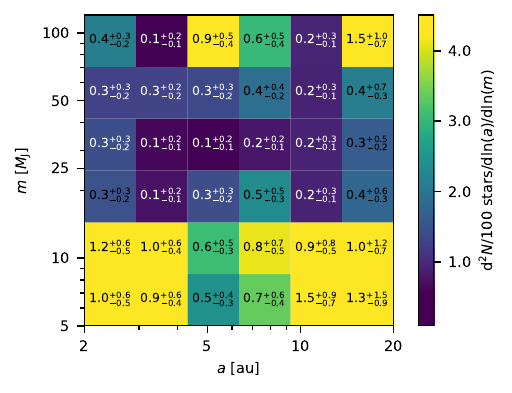}
    \caption{Occurrence rates in the $a\text{--}m$ space. The text in each cell indicates the average occurrence as a percentage (number of planets per 100 stars) along with the associated uncertainty. Text in cells with an occurrence of less than 0.3\% is displayed in white for clarity. The color of each cell indicates the density (${\mathrm{d}^2N}/(\mathrm{d}\ln a \cdot \mathrm{d}\ln m)$). The color scale is truncated at 4\% to enhance visibility.}
    \label{fig:2d-non-parametric}
\end{figure}

\section*{Analysis \& Results}\label{sec:results}
Fig.~\ref{fig:weighted-KDE} and \ref{fig:2d-non-parametric} show the occurrence rates in the parameter space of semi-major axis and companion mass.

\subsection*{Characterization of the Brown Dwarf Desert}
\begin{figure}[h!]
    \centering
    \includegraphics[width=\columnwidth]{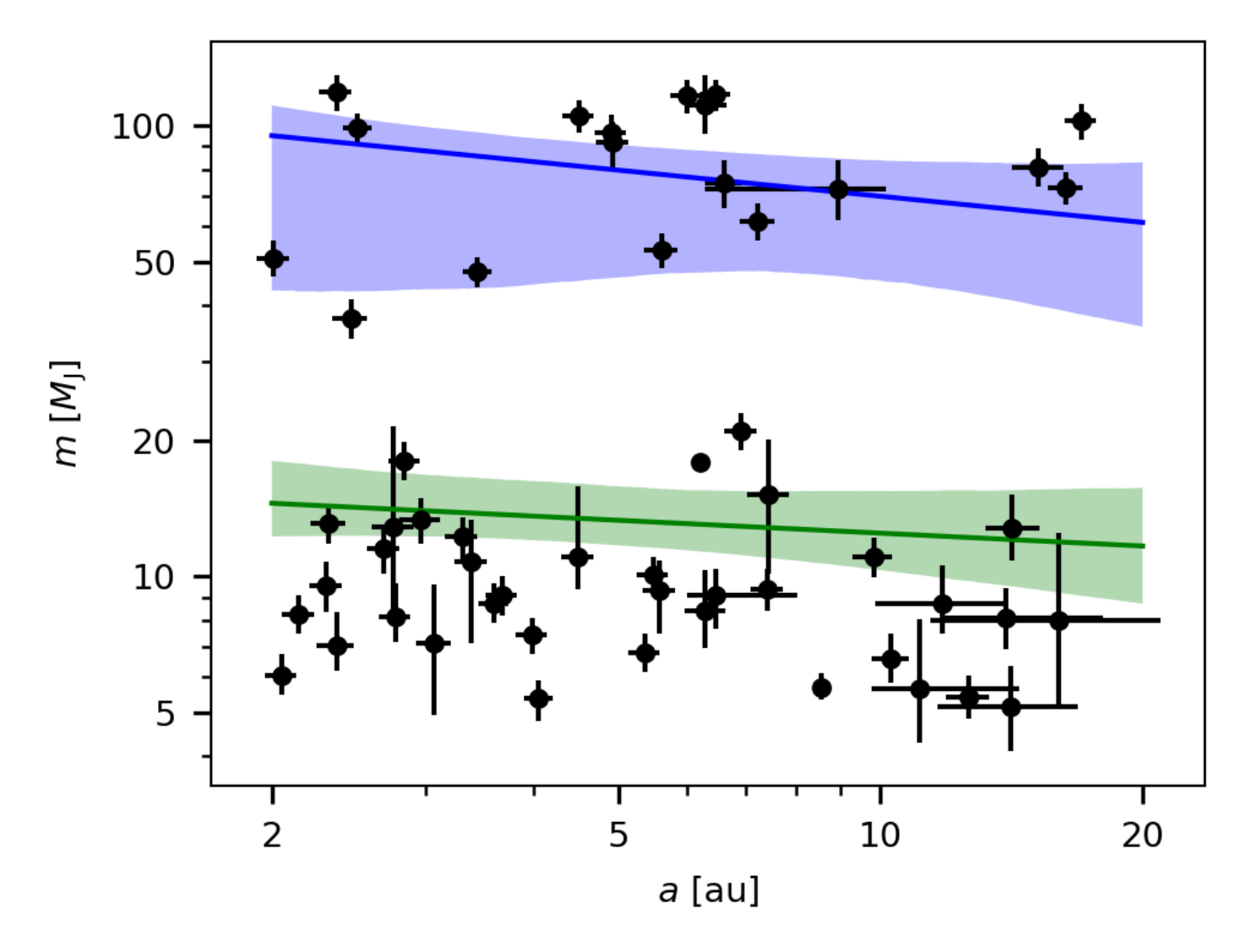}
    \caption{Boundaries of the brown dwarf desert. The upper and lower boundaries, along with their associated uncertainties, are indicated in blue and green, respectively.}
    \label{fig:upper_lower_boundaries}
\end{figure}

We observe the well-documented brown dwarf desert \cite{campbell88,gretherHowDryBrown2006,maStatisticalPropertiesBrown2014,shahafStudyMassratioDistribution2019,ungerExploringBrownDwarf2023,chenProbingShapeBrown2024} in Fig.~\ref{fig:weighted-KDE}, which is particularly evident along the mass axis. Using a combined log-normal and power-law model, we characterize the location of the desert by estimating the minimum occurrence rate {density} as a function of mass, finding $m_0 = 30.9^{+5.5}_{-4.8}\,\mj$. We also modeled the boundaries of the brown dwarf desert {with two lines: the upper and lower boundary lines are given by $\ln m = -0.2^{+0.4}_{-0.2} \ln a + 4.7^{+0.3}_{-1.0}$ and $\ln m = -0.1\pm0.2  \ln a + 2.7\pm0.3$, respectively. The median masses of the posterior distributions for the upper and lower boundary lines} are approximately $70.5^{+25.3}_{-26.6}\,\mj$ and $13.1^{+3.1}_{-2.3}\,\mj$, respectively. These values represent the median boundaries with $1\sigma$ uncertainties as shown in Fig.~\ref{fig:upper_lower_boundaries}.

To estimate the significance of the brown dwarf desert, we compare the occurrence rate {density} model by our one-dimensional model (Fig.~\ref{fig:weighted-KDE}b) with a uniform flat-line model.
{Thus, the significance of the brown dwarf desert can be quantified with PSIS-LOO. 
The ELPD our combined log-normal and power-law model is $71.5\pm4.6$, while the ELPD of the uniform model is $56.7\pm2.5$. Therefore, difference of those two models is $14.8$, indicating a $3\sigma$ significance.}

{Using the mass corresponding to the minimum occurrence rate density in the combined log-normal and power-law model, we can divide our sample into lower-mass ($m \lesssim 30\,\mj$) and higher-mass ($m \gtrsim 30\,\mj$) companion groups, respectively.} Both categories of companions exhibit a relatively constant occurrence rate density, whether above or below the desert region.
Therefore, our finding does not support an outer boundary of the brown dwarf desert up to 20\,au, suggesting that this feature may be a more universal characteristic, in contrast to previous studies that have tentatively proposed possible limits \cite{shahafStudyMassratioDistribution2019,chenProbingShapeBrown2024}. {Our results likely diverge from earlier work because, thanks to the availability of precise astrometric data, we can more reliably determine the absolute masses of companions and confidently identify brown dwarfs at wider separations. Past studies, relying chiefly on radial velocity or transit methods, were less sensitive to such long-period objects and often only provided lower mass limits. As a result, brown dwarfs at these wider orbits could have been missed or misclassified.}

\subsection*{Occurrence Rates of Giant Planets and Brown Dwarfs}
\begin{figure}
    \centering
    \includegraphics[width=\columnwidth]{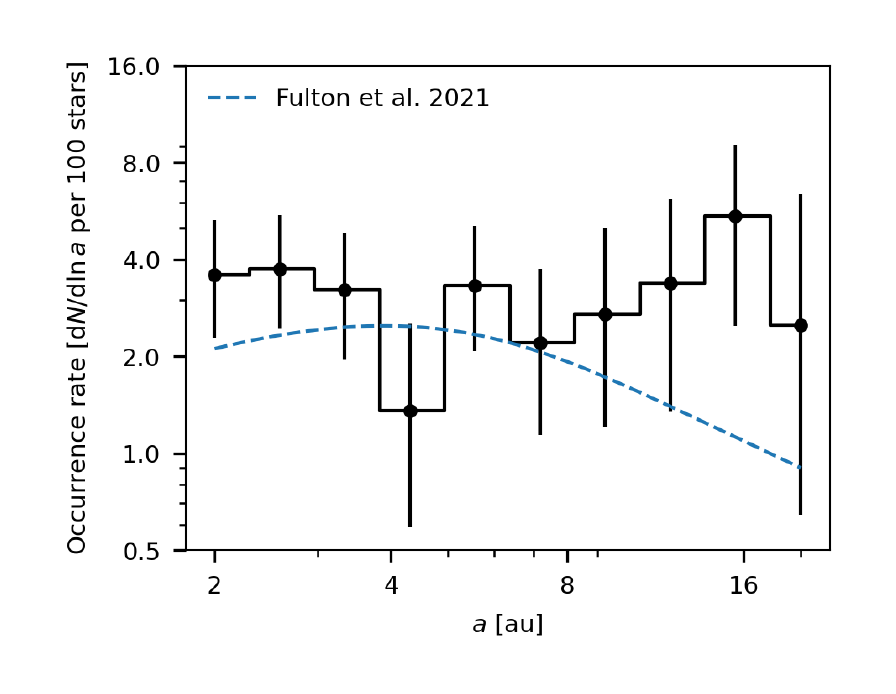}
    \caption{Occurrence rate {density} and associated errors for the mass range $5 < m < 20\,\mj$, shown by the black stepped line. Comparative data from a previous study is overlaid with a dashed blue line.}
    \label{fig:le20mj_comp}
\end{figure}

The occurrence rate as a function of mass and semi-major axis is calculated through a non-parametric analysis, achieved by dividing the $a\text{--}m$ diagram into grids. As illustrated in Fig.~\ref{fig:2d-non-parametric}, the overall occurrence rate of companions ($5< m < 120\,\mj$) within 2--20\,au is $23.1^{+2.7}_{-2.5}\%$. The occurrence rate of companions near the center of the desert (at 2--20 au and 24--42\,$\mj$) is $1.6^{+0.7}_{-0.5}$\%. In contrast, outside this boundary, giant planets (5--14\,$\mj$) have an occurrence rate of $12.6^{+2.3}_{-1.9}$\%, while stars (71--120\,$\mj$) exhibit an occurrence rate of $4.2^{+1.2}_{-1.0}$\%. Thus, the lower boundary is more evident than the upper boundary.
In particular, as shown in Fig.~\ref{fig:le20mj_comp}, we compare our occurrence rate with the result of ref. \cite{fultonCaliforniaLegacySurvey2021} within the same parameter space ($5<m<20\,\mj$ and 2 < a < 20\,au). The less massive companions may be shaped by two populations separated by a 5\,au gap. The >5\,au population is in tension with the previous prediction of a declining planet occurrence rate beyond the water-ice line \cite{fernandesHintsTurnoverSnow2019,fultonCaliforniaLegacySurvey2021}. 

\subsection*{Metallicity and Eccentricity Trends}
\begin{figure}
    \centering
    \includegraphics[width=\columnwidth]{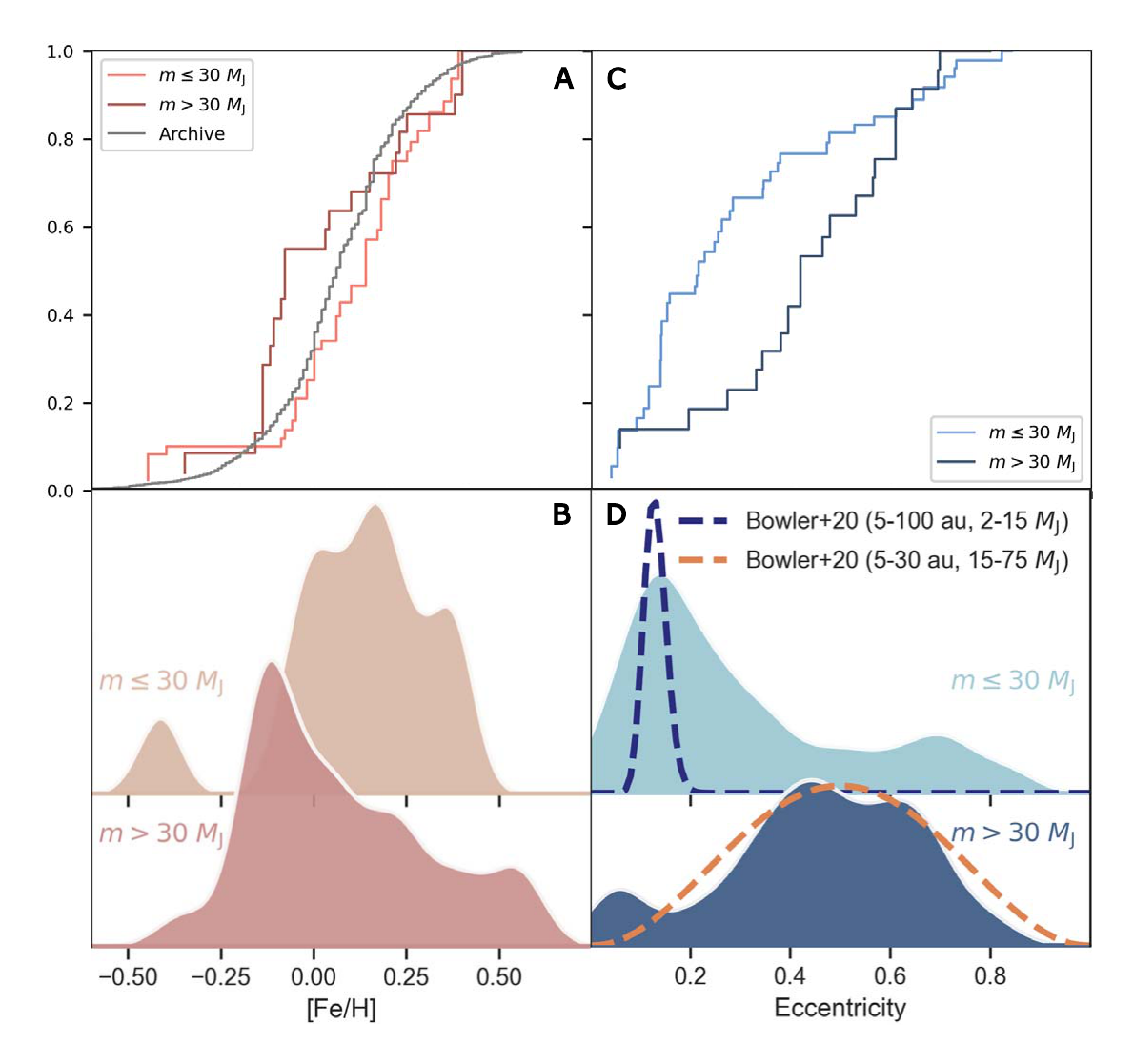}
    \caption{{Metallicity and eccentricity distributions for sample of different masses.} {(A)}, Detection efficiency-weighted cumulative distribution of host stars' metallicity and the sample of all known exoplanets collected from NASA exoplanet archive. {(B)}, Detection efficiency-weighted KDE of metallicities. 
    {(C)}, Similar to {(A)}, but for eccentricity. {(D)}, Similar to {(B)}, but for eccentricity. Selected distributions from previous results\cite{bowlerPopulationlevelEccentricityDistributions2020} are overlaid as dashed lines for reference. The y-axis of {(B)} and {(D)} are scaled.}
    \label{fig:feh-ecc-dist}
\end{figure}

{The brown dwarf desert and the potential new population beyond 5\,au are further investigated through their eccentricity and metallicity distributions.}
As depicted in Figure~\ref{fig:feh-ecc-dist}a and b, companions with $m \leq 30\,\mj$ do not exhibit a significant preference for metallicity when compared to companions with $m > 30\,\mj$. However, among objects with near-solar metallicity (-0.25 -- 0.25), the more massive companions tend to be metal-poor relative to the lower mass companions, as well as all known exoplanets cataloged in the NASA Exoplanet Archive\footnote{\url{https://exoplanetarchive.ipac.caltech.edu/}} (Kolmogorov-Smirnov test p-values: 0.01, 0.02 respectively). Although NASA Exoplanet Archive is a heterogeneous sample, we present it here solely as a reference.
Previous studies have indicated that brown dwarfs and low-mass stars generally exhibit lower metallicities than giant planets, based on transit and radial velocity samples \cite{jenkinsObservedDistributionSpectroscopic2015,schlaufmanEvidenceUpperBound2018}, which may suggest different formation mechanisms \cite{santosSpectroscopicFe982004,sozzettiKeckHIRESDoppler2006}. Our findings support this weak correlation, particularly when we rigorously account for bias and employ a stringent occurrence rate calculation. 

\begin{figure}
    \centering
    \includegraphics[width=\columnwidth]{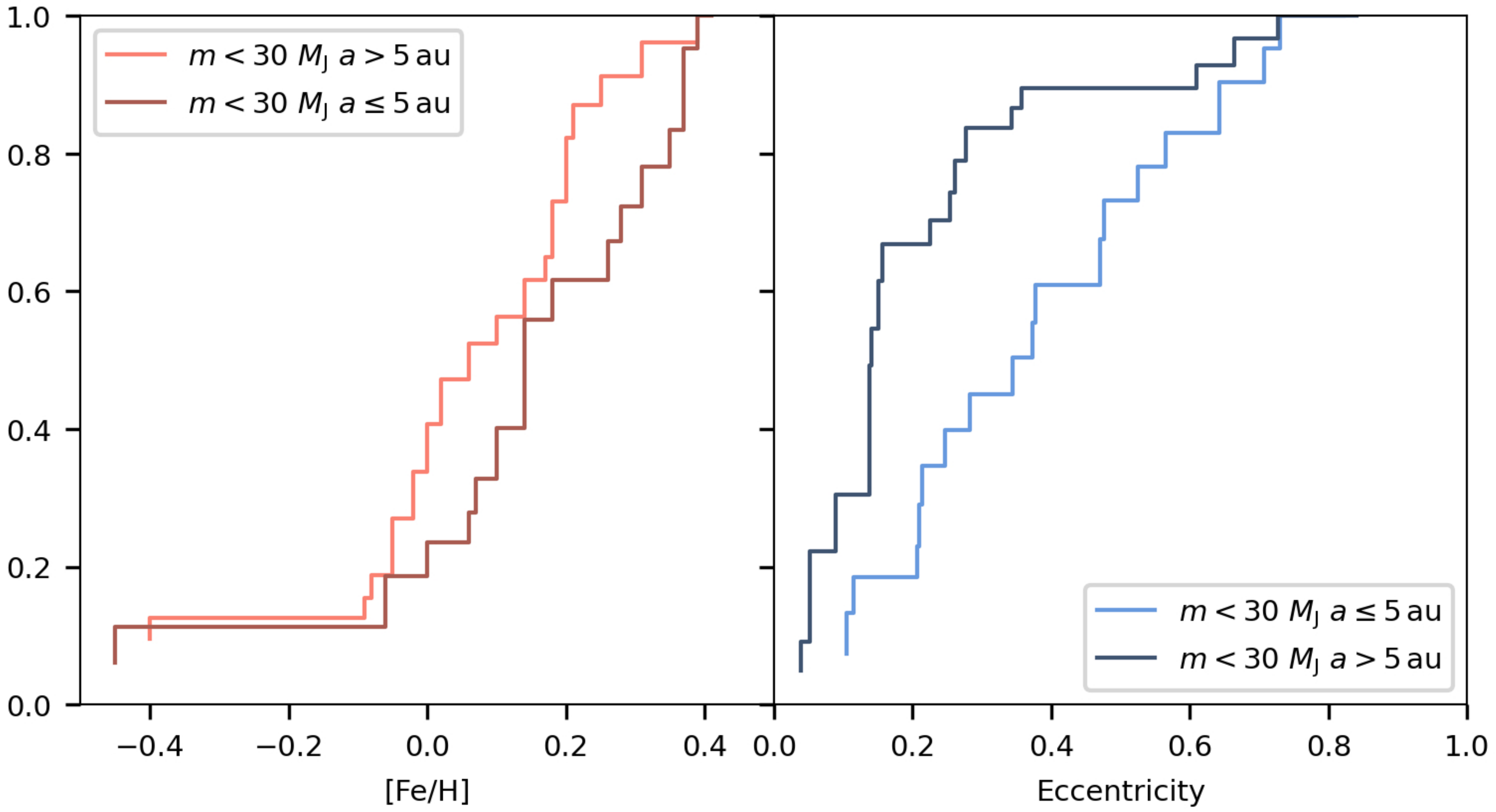}
    \caption{Cumulative distributions of metallicities and eccentricities for the inner ($\leq 5\,$au) and outer ($> 5\,$au) low mass companions ($m < 30\,\mj$).}
    \label{fig:cdf-ecc-feh-le30mj}
\end{figure}

Furthermore, we examine the origins of distinct populations by analyzing their eccentricity distributions while accounting for detection efficiency. As depicted in Fig.~\ref{fig:feh-ecc-dist}c and d, both cumulative and density distributions show less massive companions exhibit lower eccentricities compared to massive ones.
The eccentricities of our lower-mass companions are concentrated at low values, similar to those of giant planets detected through direct imaging \citep[5--100\,au, 2--15\,$\mj$;][]{bowlerPopulationlevelEccentricityDistributions2020}. Likewise, the higher-mass companions exhibit eccentricities that align well with those of the brown dwarf sample \citep[5--30\,au, 15--75\,$\mj$;][]{bowlerPopulationlevelEccentricityDistributions2020}. The notably discrepant eccentricities of different masses suggest that they may have different formation or evolution histories. Also, the differences in eccentricities between the inner ($\leq 5\,$au) and outer ($> 5\,$au) low mass companions may hint that they are, in fact, two groups originating from distinct sources (Fig.~\ref{fig:cdf-ecc-feh-le30mj}).

\section*{Discussion}\label{sec:discussion}
Considering that lower mass companions are enriched in metals and lower eccentricities than the higher mass companions, it is likely that they are formed through different channels. Generally, giant planets and brown dwarfs form either by core accretion \citep[CA;][]{perriHydrodynamicInstabilitySolar1974,mizunoFormationGiantPlanets1980,pollackFormationGiantPlanets1996} from colliding via agglomeration of planetesimals and subsequent gas capture \cite{johansenFormingPlanetsPebble2017,liuTalePlanetFormation2020} or by direct collapse in the disc via gravitational instability  \citep[GI;][]{cameronEvolutionGiantGaseous1982,boss97,mayerFormationGiantPlanets2002,kratterGravitationalInstabilitiesCircumstellar2016}. CA is unlikely to form high-mass ($\gtrsim 30\,\mj$) companions on wide orbit due to long accretion time scale \cite{idaDeterministicModelPlanetary2004,johansenExploringConditionsForming2019,emsenhuberNewGenerationPlanetary2021a}, while GI is thought to be unlikely to form low-mass ($\lesssim 30\,\mj$) companions on close-in orbit due to inefficient cooling \cite{Rafikov05,durisenDiskHydrodynamics2011}.
{Many hydrodynamical simulations \citep{vorobyovBurstModeProtostellar2006,VB10,vorobyovVARIABLEPROTOStelLARACCRETION2015,BoleyEtal10,InutsukaEtal09,BaruteauEtal11,chaNumericalSimulationSuperEarth2011,MichaelEtal11,zhu12} also indicate that planets formed via GI can migrate inward very efficiently, resulting in the formation of close-in ($\sim$10\,au) giant planets.}

\subsection*{Model Comparisons: Hybrid Formation Hypothesis}

\begin{figure}
    \centering
    \includegraphics[width=\columnwidth]{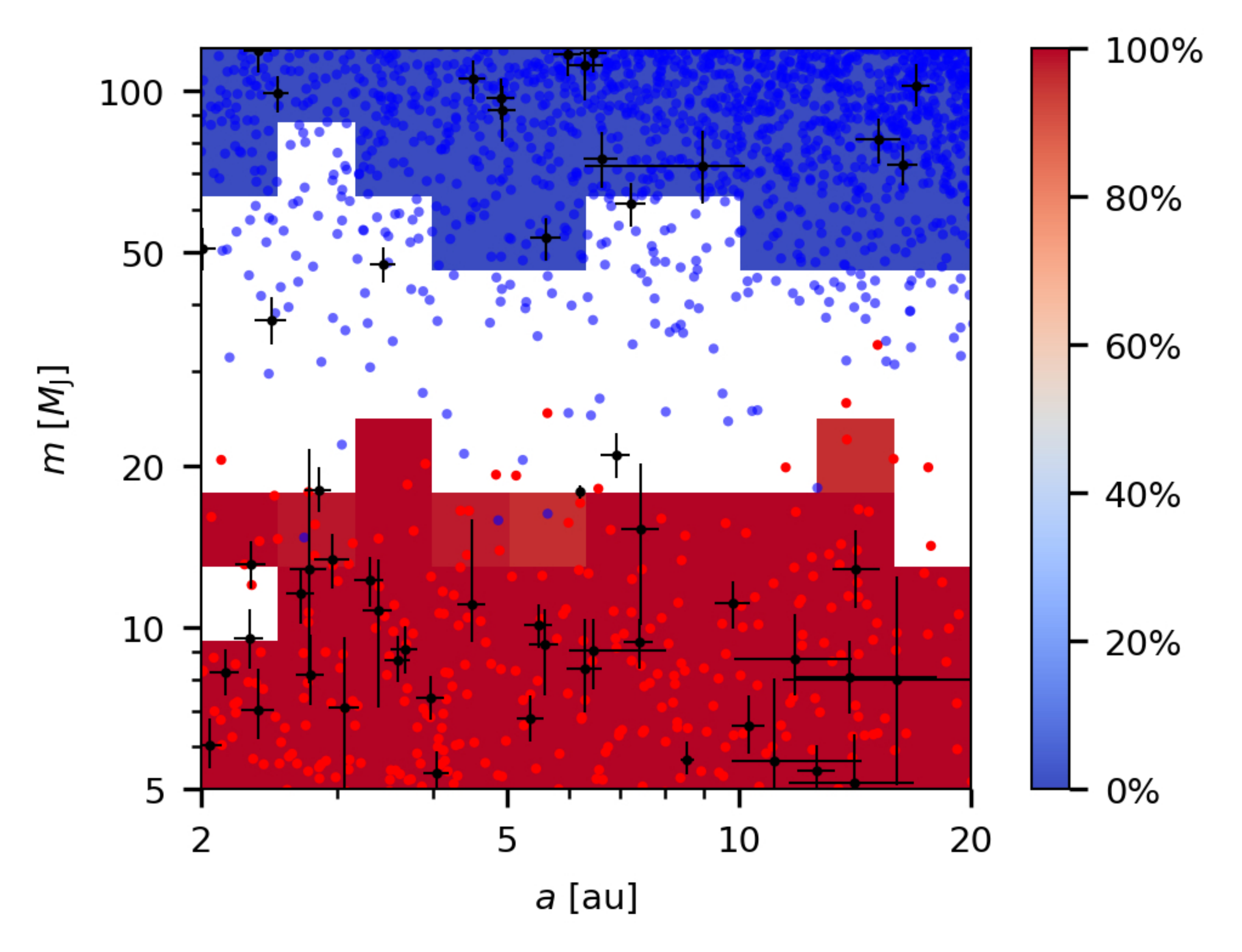}
    \caption{Results of Population Synthesis. Fractional occurrence rate of CA based on combined population synthesis model. The color scale represents the proportion of CA occurrence rate in relation to the total occurrence rate for both CA and GI formation mechanisms. Occurrence rate densities below 0.5\% are truncated to illustrate the brown dwarf desert more clearly. Populations associated with CA and GI are represented by red and blue points, respectively, while observational data, including uncertainties, are shown as black points.}
    \label{fig:boundary}
\end{figure}

To investigate the formation pathways of the distinct populations in our sample, we performed hybrid population synthesis simulations that combine both CA and GI. 
{Our GI model is functionally equivalent to the low-mass end of the stellar formation pathway, enabling it to form the observed close, low-mass stars.}
The combined model reproduces the observed distributions more accurately than either formation mechanism alone (see Fig.~\ref{fig:boundary}). The fractional occurrence rates of CA and GI shown in Fig.~\ref{fig:boundary} indicate that the observed sample is consistent with a mixture of both processes, with the brown dwarf desert emerging naturally as the boundary between the two formation regimes.

\subsection*{Implications for Giant Planet and Brown Dwarf Formation}

Interestingly, {GI} models have demonstrated a desert at various masses, for example, a reduction in frequency at $\sim 50\,\mj$ demonstrated in population synthesis models \cite{forganPopulationSynthesisModel2018}, and a break at much lower masses in hydrodynamical models \cite{HallCEtal17}. However, neither the population synthesis models nor the hydrodynamical models are able to explain our observed eccentricity distribution, which requires a combination of both CA and GI. To elucidate the observed features and gain a comprehensive understanding of the formation processes involved, further observations and more advanced models are required.

\section*{Conclusions}\label{sec:conclusions}
In this study, we measured the occurrence rates of giant planets, brown dwarfs, and stellar companions ($5$--$120\,\mj$) in wide orbits (2--20\,au), identified through combined radial velocity and astrometric observations. Our main findings are summarized below:

\begin{enumerate}
    \item The minimum of the brown dwarf desert is located at $30.9^{+5.5}_{-4.8}\,\mj$. The {median mass of} upper boundary is $70.5^{+25.3}_{-26.6}\,\mj$, while the {median mass of} lower boundary is $13.1^{+3.1}_{-2.3}\,\mj$. No outer boundary is detected within 20\,au in our sample.

    \item The occurrence rate of our giant planets ($5$--$14\,\mj$) is $12.6^{+2.3}_{-1.9}\%$, while stars ($71$--$120\,\mj$) show an occurrence rate of $4.2^{+1.2}_{-1.0}\%$, both within a semi-major axis range of 2--20\,au. We also find a population of giant planets emerging beyond 5\,au, in contradiction to previous results.

    \item High-mass companions (> 30\,\mj) exhibit relatively lower metallicity and higher eccentricity than low-mass companions (< 30\,\mj), suggesting different formation channels. Additionally, inner and outer low-mass companions display distinct eccentricity distributions, indicating that they may originate from two separate sources.

    \item Our hybrid population synthesis results support the existence of the brown dwarf desert due to the formation boundary between CA and GI.

\end{enumerate}

{Because the brown dwarf desert may depend sensitively on host-star properties (e.g., spectral type, metallicity), more advanced population synthesis studies should explicitly incorporate these dependencies. Advancing this work will require larger, more uniform samples and additional observations to delineate formation boundaries in this uncertain regime and to robustly quantify the occurrence and properties of giant planets and brown dwarfs.}

\acknow{{We thank the referees for their constructive suggestions, which have greatly improved this paper.} We thank S. Ida, Y. Hori, D. N. C. Lin, A. Vigan, K. Rice, H. Deng, H. Jones, and J. Jenkins for their efforts and helpful discussions. This work is supported by the National Key R\&D Program of China, {No. 2024YFA1611801 and No. 2024YFC2207700}, by the National Natural Science Foundation of China (NSFC) under Grant No. 12473066, {and No. 12273023}, by Shanghai Jiao Tong University 2030 Initiative, Science and Technology Commission of Shanghai Municipality (project No. 23JC1410200) and Zhangjiang National Innovation Demonstration Zone (project No. ZJ2023-ZD-003). The collaboration of this work is partially supported by the SJTU Global Strategic Partnership Fund (2022SJTU-Warwick).} 

\showacknow{} 



\section*{References}\label{sec:references}
\bibliography{my_boundary2,nm}

\newpage
\onecolumn
\hypertarget{si}{}
\setcounter{figure}{0}
\setcounter{table}{0}
\renewcommand{\thefigure}{S\arabic{figure}}
\renewcommand{\thetable}{S\arabic{table}}

\begin{center}
\includegraphics[width=9.95cm]{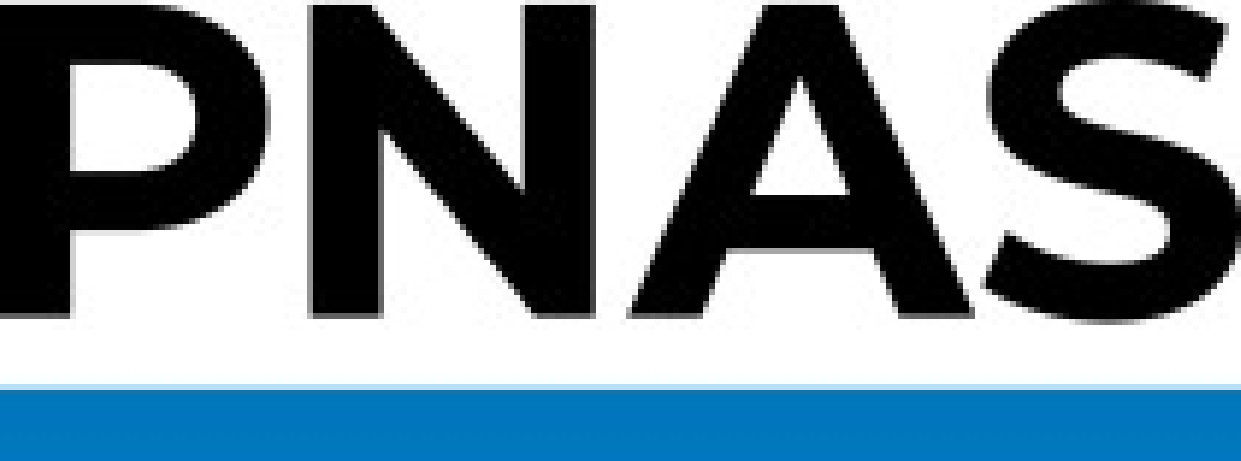}
\end{center}

\vskip45pt
\begingroup
\raggedright
{\Huge\sffamily\bfseries Supporting Information for\par} 
\bigskip
{\LARGE\sffamily\bfseries A universal brown dwarf desert formed between planets and stars \par}
\bigskip
{\sffamily\bfseries Kaiming Cui, Guangyao Xiao, Fabo Feng, Beibei Liu, Sergei Nayakshin, Cassandra Hall, Kangrou Guo, Dong Lai,
Masahiro Ogihara, Yicheng Rui, Alan P. Boss, R. Paul Butler, and Yifan Xuan \par\bigskip Fabo Feng \par Email: ffeng@sjtu.edu.cn \par}
\endgroup
\bigskip
\section*{This PDF file includes:}
\begin{list}{}{%
\setlength\leftmargin{2em}%
\setlength\itemsep{0pt}%
\setlength\parsep{0pt}}
\item Supporting text
\item Figs.~S1 to S19
\item Tables S1 to S4
\item SI References
\end{list}

\newpage
\onecolumn

\section*{Sample selection detail}
We list the four RV surveys we are using as below:
\begin{itemize}
\item {HARPS GTO. The HARPS high-precision GTO project aims to detect very low-mass exoplanets by improving radial velocity measurement accuracy below 1 $\rm m\,s^{-1}$. This sample consists of $\sim1000$ FGK stars observed in the context of three HARPS-GTO planet search programs \cite{mayor03,Lo-Curto2010,Santos2011A&A...526A.112S}. The stars were selected from a well-defined volume in the solar neighborhood. These stars are mostly slow rotators, non-evolved FGK-type dwarfs, and have low chromospheric activity levels, which are suitable for radial velocity surveys.}

\item {California Planet Search. The CPS team \cite{Howard2010ApJ...721.1467H} has conducted several RV surveys over the past three decades to detect exoplanets and study their intrinsic occurrence. They constructed stellar samples without bias toward stars with known planets, or an increased likelihood of hosting planets (e.g., metal-rich systems). The primary surveys include the Keck Planet Search \cite{cummingKeckPlanetSearch2008} of 585 FGKM stars using HIRES, the Eta-Earth Survey \cite{Howard2010Sci...330..653H} of 166 Sunlike stars, the APF-50 survey (Automated Planet Finder \cite{Radovan2014SPIE.9145E..2BR}) of 50 bright stars, and the Lick Planet Search \cite{Fischer2014ApJS..210....5F} of 387 dwarfs. More detains about the observations of CPS can be found in ref\cite{rosenthalCaliforniaLegacySurvey2021}. Following ref. \cite{rosenthalCaliforniaLegacySurvey2021}, we excluded stars that were deliberately chosen because they were known to host planets, were metal-rich, or were evolved. For instance, the ``N2K'' and ``M2K'' programs specifically targeted metal-rich stars in the search for gas giants \cite{fischerPlanetMetallicityCorrelation2005, Apps2010PASP..122..156A}. However, we will include those that has been observed as part of HARPS GTO. Finally, all follow-up observations of transiting systems were also excluded.} 

\item {Anglo-Australian Planet Search. The program began operations in January 1998, using the 3.9-m Anglo-Australian Telescope equipped with the University College London Echelle Spectrograph (UCLES). For the first 14 years, AAPS monitored about 250 solar-type stars. Since 2013, it has refined its target list to approximately 120 stars most suitable for detecting Jupiter analogs, due to limited telescope time. The AAPS catalogue is listed in ref. \cite{Jones2002MNRAS.337.1170J, Tinney2003ApJ...587..423T}, and further details of the observing program are described in ref.\cite{butler01}}

\item {Magellan Planet Search Program. The Program has been monitoring more than 690 G, F, and K main-sequence stars since late 2002 \cite{Arriagada2011ApJ...734...70A}, using high-resolution echelle spectrographs at Las Campanas Observatory. Initially, observations were primarily conducted with the MIKE spectrograph \cite{Bernstein2003SPIE.4841.1694B} mounted on the 6.5 m Clay Telescope (Magellan II). In 2010, the survey transitioned to the Carnegie Planet Finder Spectrograph \cite[PFS;][]{Crane2006SPIE.6269E..31C,Crane2008SPIE.7014E..79C,Crane2010SPIE.7735E..53C}, a temperature-controlled, high-resolution echelle spectrograph designed for precision RV measurements, on Magellan II. We also exclude stars that observed by Magellan-Tess Survey \citep[MTS;][]{Teske2021ApJS..256...33T}.}
\end{itemize}


{The stars that are only from the planet survey program of SOPHIE or ELODIE are not considered, as their complete volume-limited sample are not available, and most of their currently published RVs are known to possess variations induced by substellars \cite{Kiefer2019A&A...631A.125K}. This will introduce additional selection bias to our stellar selection.}

{Then we further restrict our sample using the following selection criteria:}
\begin{itemize}
\item {the number and duration of observations in the sample are larger than 5 RVs and 1 yr.}

\item The stellar mass should be in the range of $0.6 \lesssim M_{\star} \lesssim 1.4 \,M_{\odot}$ to ensure the selection of late FGK-type mass range\cite{pecautINTRINSICCOLORSTEMPERATURES2013,stassun19}.

\item Giant stars that have evolved from the main sequence are excluded {using an approximate threshold of ${\rm log}\,g < 3.9$\,dex\cite{Huber2016ApJS}.}

\end{itemize}

\begin{figure}[h!]
    \centering
    \includegraphics[width=0.5\columnwidth]{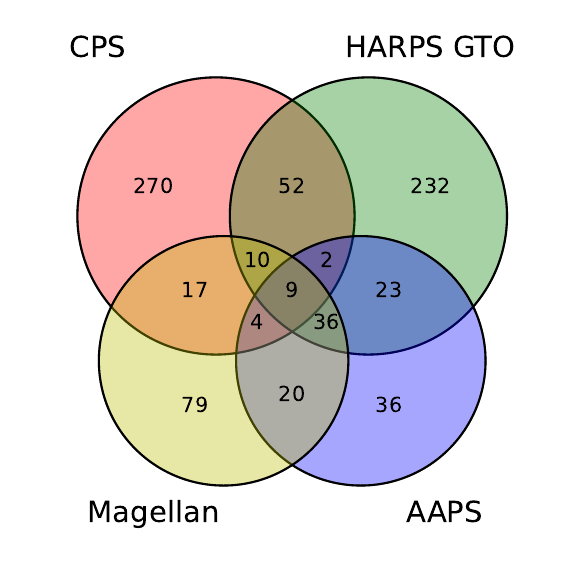}
    \caption{The number of cross-matched stars for HARPS GTO, CPS, AAPS and the Magellan Planet Search Program in our sample. Their overlaps are quite apparent and complicated. }
    \label{fig:all-sample}
\end{figure}

Fig.~\ref{fig:all-sample} shows our entire sample and illustrates the overlap between these four surveys. Our companions are listed in Table~\ref{tab:sample}.

\begin{longtable}{cccccccccc}
\caption{Catalog of companions}
\label{tab:sample}\\
\hline
Host & Ra & Dec & $m\,(\mj)$ & $q$ & $e$ & $a$\,(au) & $P$\,(yr) & $I$ & $\Omega$ \\
\hline
HD151450 & 252.01 & -15.96 & $5.14^{+1.19}_{-1.05}$ & 4.3E-03$^\mathrm{+1.2E-03}_\mathrm{-1.1E-03}$ & $0.14^{+0.11}_{-0.08}$ & $14.10^{+2.77}_{-2.49}$ & $49.68^{+14.66}_{-11.91}$ & $42.37^{+11.09}_{-8.86}$ & $39.72^{+15.22}_{-18.93}$ \\
HD53665 & 106.47 & -1.02 & $5.64^{+2.40}_{-1.36}$ & 4.5E-03$^\mathrm{+2.0E-03}_\mathrm{-1.3E-03}$ & $0.23^{+0.04}_{-0.10}$ & $11.10^{+3.32}_{-1.33}$ & $28.85^{+20.31}_{-0.50}$ & $54.52^{+2.25}_{-19.88}$ & $62.09^{+60.29}_{-71.82}$ \\
HIP35965 & 111.20 & -51.33 & $8.02^{+4.46}_{-2.92}$ & 9.8E-03$^\mathrm{+5.6E-03}_\mathrm{-3.8E-03}$ & $0.14^{+0.11}_{-0.09}$ & $16.02^{+4.94}_{-4.62}$ & $71.53^{+37.27}_{-27.85}$ & $43.70^{+8.53}_{-8.32}$ & $160.10^{+18.84}_{-17.54}$ \\
HD175167 & 285.00 & -69.94 & $9.57^{+1.24}_{-1.21}$ & 9.1E-03$^\mathrm{+1.7E-03}_\mathrm{-1.7E-03}$ & $0.53^{+0.00}_{-0.00}$ & $2.31^{+0.09}_{-0.10}$ & $3.49^{+0.00}_{-0.00}$ & $42.18^{+5.65}_{-4.11}$ & $56.60^{+6.54}_{-6.25}$ \\
HD217786 & 345.78 & -0.43 & $37.43^{+3.82}_{-3.78}$ & 3.1E-02$^\mathrm{+5.4E-03}_\mathrm{-5.4E-03}$ & $0.33^{+0.00}_{-0.00}$ & $2.47^{+0.11}_{-0.12}$ & $3.57^{+0.00}_{-0.00}$ & $21.08^{+0.96}_{-1.05}$ & $174.11^{+5.63}_{-4.64}$ \\
HD39213 & 87.32 & -37.51 & $118.94^{+10.81}_{-10.98}$ & 1.2E-01$^\mathrm{+2.0E-02}_\mathrm{-2.0E-02}$ & $0.19^{+0.01}_{-0.01}$ & $2.37^{+0.10}_{-0.11}$ & $3.56^{+0.01}_{-0.01}$ & $144.39^{+1.05}_{-1.20}$ & $85.31^{+6.67}_{-6.65}$ \\
HD72659 & 128.51 & -1.57 & $12.82^{+2.34}_{-1.97}$ & 1.1E-02$^\mathrm{+2.3E-03}_\mathrm{-2.3E-03}$ & $0.05^{+0.04}_{-0.03}$ & $14.17^{+1.07}_{-0.97}$ & $50.83^{+4.85}_{-3.69}$ & $19.62^{+5.20}_{-3.36}$ & $168.95^{+10.40}_{-15.57}$ \\
HIP48953 & 149.80 & -9.18 & $98.93^{+7.65}_{-7.73}$ & 1.3E-01$^\mathrm{+1.9E-02}_\mathrm{-1.9E-02}$ & $0.39^{+0.01}_{-0.01}$ & $2.51^{+0.09}_{-0.10}$ & $4.43^{+0.02}_{-0.02}$ & $100.38^{+2.71}_{-3.01}$ & $255.42^{+1.29}_{-1.25}$ \\
HD28185 & 66.61 & -10.55 & $5.68^{+0.44}_{-0.36}$ & 5.5E-03$^\mathrm{+8.0E-04}_\mathrm{-8.0E-04}$ & $0.14^{+0.02}_{-0.03}$ & $8.54^{+0.21}_{-0.14}$ & $25.27^{+0.91}_{-0.61}$ & $73.00^{+10.00}_{-8.40}$ & $287.00^{+19.00}_{-22.00}$ \\
HD62364 & 113.91 & -75.54 & $17.90^{+0.54}_{-0.53}$ & 1.4E-02$^\mathrm{+2.0E-03}_\mathrm{-2.0E-03}$ & $0.61^{+0.00}_{-0.00}$ & $6.21^{+0.07}_{-0.07}$ & $14.13^{+0.06}_{-0.06}$ & $133.10^{+1.30}_{-1.30}$ & $278.40^{+3.00}_{-3.00}$ \\
HD106515A & 183.78 & -7.26 & $11.06^{+4.84}_{-1.67}$ & 1.1E-02$^\mathrm{+5.2E-03}_\mathrm{-2.2E-03}$ & $0.56^{+0.02}_{-0.03}$ & $4.50^{+0.18}_{-0.19}$ & $9.82^{+0.16}_{-0.09}$ & $88.92^{+42.18}_{-39.90}$ & $283.38^{+79.73}_{-74.41}$ \\
HD10697 & 26.23 & 20.08 & $6.03^{+0.73}_{-0.57}$ & 5.8E-03$^\mathrm{+9.9E-04}_\mathrm{-9.0E-04}$ & $0.10^{+0.01}_{-0.01}$ & $2.05^{+0.08}_{-0.09}$ & $2.94^{+0.00}_{-0.00}$ & $86.12^{+19.96}_{-20.53}$ & $38.85^{+15.08}_{-21.59}$ \\
HD111232 & 192.22 & -68.42 & $8.26^{+0.83}_{-0.78}$ & 8.2E-03$^\mathrm{+1.3E-03}_\mathrm{-1.3E-03}$ & $0.21^{+0.00}_{-0.00}$ & $2.15^{+0.09}_{-0.10}$ & $3.20^{+0.00}_{-0.00}$ & $93.52^{+16.62}_{-18.06}$ & $358.31^{+52.53}_{-32.10}$ \\
HD125612 & 215.22 & -17.48 & $7.40^{+0.71}_{-0.67}$ & 6.7E-03$^\mathrm{+1.1E-03}_\mathrm{-1.0E-03}$ & $0.11^{+0.01}_{-0.01}$ & $3.98^{+0.16}_{-0.17}$ & $7.73^{+0.04}_{-0.04}$ & $88.20^{+16.31}_{-15.98}$ & $267.00^{+22.67}_{-22.90}$ \\
HD126614 & 216.70 & -5.18 & $81.04^{+7.87}_{-7.83}$ & 8.1E-02$^\mathrm{+1.3E-02}_\mathrm{-1.3E-02}$ & $0.06^{+0.02}_{-0.02}$ & $15.17^{+1.04}_{-1.03}$ & $59.98^{+4.70}_{-5.06}$ & $16.29^{+0.88}_{-0.77}$ & $284.66^{+2.60}_{-3.30}$ \\
HD127124 & 217.82 & -51.38 & $111.43^{+17.79}_{-15.42}$ & 1.3E-01$^\mathrm{+2.5E-02}_\mathrm{-2.3E-02}$ & $0.56^{+0.04}_{-0.03}$ & $6.29^{+0.36}_{-0.32}$ & $16.96^{+1.13}_{-0.69}$ & $165.10^{+0.75}_{-0.80}$ & $185.78^{+15.05}_{-16.59}$ \\
HD127506 & 217.69 & 35.45 & $47.52^{+3.59}_{-3.64}$ & 6.2E-02$^\mathrm{+8.5E-03}_\mathrm{-8.5E-03}$ & $0.70^{+0.00}_{-0.00}$ & $3.45^{+0.13}_{-0.14}$ & $7.34^{+0.10}_{-0.14}$ & $128.92^{+1.47}_{-0.63}$ & $223.41^{+1.34}_{-1.76}$ \\
HD136118 & 229.73 & -1.59 & $13.15^{+1.31}_{-1.31}$ & 1.1E-02$^\mathrm{+1.8E-03}_\mathrm{-1.8E-03}$ & $0.34^{+0.01}_{-0.01}$ & $2.33^{+0.10}_{-0.11}$ & $3.25^{+0.01}_{-0.01}$ & $116.78^{+4.02}_{-5.87}$ & $106.37^{+14.01}_{-9.11}$ \\
HD142 & 1.58 & -49.08 & $11.08^{+1.10}_{-1.12}$ & 8.7E-03$^\mathrm{+1.5E-03}_\mathrm{-1.5E-03}$ & $0.28^{+0.03}_{-0.03}$ & $9.82^{+0.51}_{-0.53}$ & $27.82^{+1.05}_{-0.88}$ & $90.37^{+10.50}_{-12.32}$ & $181.93^{+3.21}_{-3.55}$ \\
HD145675 & 242.60 & 43.82 & $8.15^{+1.54}_{-0.98}$ & 8.5E-03$^\mathrm{+1.9E-03}_\mathrm{-1.5E-03}$ & $0.37^{+0.01}_{-0.01}$ & $2.77^{+0.11}_{-0.12}$ & $4.83^{+0.00}_{-0.01}$ & $144.65^{+6.28}_{-3.24}$ & $222.31^{+6.42}_{-6.11}$ \\
HD150554 & 250.23 & 21.95 & $92.10^{+9.95}_{-11.91}$ & 7.9E-02$^\mathrm{+1.3E-02}_\mathrm{-1.4E-02}$ & $0.57^{+0.00}_{-0.00}$ & $4.93^{+0.20}_{-0.21}$ & $10.10^{+0.00}_{-0.00}$ & $133.10^{+1.71}_{-14.36}$ & $236.98^{+2.64}_{-30.71}$ \\
HD156279 & 258.10 & 63.35 & $10.09^{+0.98}_{-0.95}$ & 1.0E-02$^\mathrm{+1.6E-03}_\mathrm{-1.6E-03}$ & $0.26^{+0.01}_{-0.01}$ & $5.48^{+0.22}_{-0.24}$ & $13.19^{+0.12}_{-0.12}$ & $74.70^{+37.00}_{-10.53}$ & $288.95^{+12.74}_{-37.24}$ \\
HD16160 & 39.03 & 6.89 & $102.37^{+9.74}_{-9.10}$ & 1.3E-01$^\mathrm{+1.9E-02}_\mathrm{-1.8E-02}$ & $0.64^{+0.00}_{-0.00}$ & $17.00^{+0.68}_{-0.72}$ & $76.07^{+1.86}_{-1.83}$ & $45.78^{+2.58}_{-3.71}$ & $121.68^{+4.98}_{-2.70}$ \\
HD167665 & 274.35 & -28.29 & $53.12^{+4.70}_{-4.82}$ & 4.4E-02$^\mathrm{+7.1E-03}_\mathrm{-7.1E-03}$ & $0.34^{+0.01}_{-0.01}$ & $5.61^{+0.24}_{-0.26}$ & $12.15^{+0.03}_{-0.04}$ & $95.20^{+7.47}_{-9.86}$ & $219.02^{+1.58}_{-1.41}$ \\
HD168443 & 275.02 & -9.60 & $18.04^{+1.82}_{-1.64}$ & 1.8E-02$^\mathrm{+2.8E-03}_\mathrm{-2.7E-03}$ & $0.21^{+0.00}_{-0.00}$ & $2.84^{+0.11}_{-0.12}$ & $4.80^{+0.00}_{-0.00}$ & $91.22^{+22.28}_{-16.09}$ & $121.41^{+16.76}_{-16.58}$ \\
HD16905 & 40.01 & -61.35 & $9.08^{+1.30}_{-1.43}$ & 1.1E-02$^\mathrm{+2.0E-03}_\mathrm{-2.2E-03}$ & $0.66^{+0.06}_{-0.07}$ & $6.47^{+1.56}_{-0.47}$ & $18.36^{+7.61}_{-1.49}$ & $19.66^{+3.33}_{-1.95}$ & $111.90^{+11.97}_{-14.51}$ \\
HD169830 & 276.96 & -29.82 & $7.11^{+2.50}_{-2.20}$ & 5.8E-03$^\mathrm{+2.2E-03}_\mathrm{-2.0E-03}$ & $0.25^{+0.02}_{-0.02}$ & $3.07^{+0.13}_{-0.14}$ & $4.98^{+0.02}_{-0.02}$ & $24.47^{+12.74}_{-7.21}$ & $16.52^{+25.03}_{-30.71}$ \\
HD176535 & 285.33 & -13.69 & $72.57^{+11.51}_{-10.86}$ & 9.1E-02$^\mathrm{+1.8E-02}_\mathrm{-1.8E-02}$ & $0.42^{+0.04}_{-0.13}$ & $8.95^{+1.21}_{-2.66}$ & $30.89^{+4.77}_{-13.29}$ & $144.31^{+10.84}_{-8.71}$ & $139.10^{+66.98}_{-5.04}$ \\
HD181234 & 290.00 & -9.32 & $9.38^{+1.01}_{-0.99}$ & 9.5E-03$^\mathrm{+1.6E-03}_\mathrm{-1.6E-03}$ & $0.73^{+0.01}_{-0.01}$ & $7.42^{+0.31}_{-0.33}$ & $20.79^{+0.31}_{-0.33}$ & $98.73^{+24.47}_{-41.61}$ & $346.55^{+36.81}_{-20.20}$ \\
HD183263 & 292.10 & 8.36 & $9.31^{+1.52}_{-1.82}$ & 8.4E-03$^\mathrm{+1.7E-03}_\mathrm{-2.0E-03}$ & $0.05^{+0.01}_{-0.01}$ & $5.58^{+0.23}_{-0.25}$ & $12.78^{+0.05}_{-0.05}$ & $73.72^{+63.94}_{-32.08}$ & $284.04^{+21.50}_{-24.83}$ \\
HD185414 & 293.98 & 56.98 & $117.47^{+9.16}_{-9.43}$ & 1.1E-01$^\mathrm{+1.5E-02}_\mathrm{-1.5E-02}$ & $0.69^{+0.00}_{-0.00}$ & $6.47^{+0.25}_{-0.27}$ & $15.17^{+0.17}_{-0.05}$ & $93.14^{+1.92}_{-2.23}$ & $108.82^{+1.42}_{-0.89}$ \\
HD186265 & 296.10 & -26.95 & $116.70^{+9.89}_{-10.13}$ & 1.1E-01$^\mathrm{+1.8E-02}_\mathrm{-1.8E-02}$ & $0.48^{+0.00}_{-0.00}$ & $5.99^{+0.24}_{-0.26}$ & $14.19^{+0.06}_{-0.05}$ & $63.16^{+2.13}_{-1.83}$ & $160.60^{+1.63}_{-1.49}$ \\
HD190406 & 301.02 & 17.07 & $72.92^{+6.01}_{-6.20}$ & 6.4E-02$^\mathrm{+9.8E-03}_\mathrm{-9.8E-03}$ & $0.46^{+0.01}_{-0.01}$ & $16.34^{+0.71}_{-0.75}$ & $61.63^{+1.54}_{-1.50}$ & $91.15^{+4.79}_{-5.85}$ & $146.72^{+3.18}_{-3.88}$ \\
HD204313 & 322.05 & -21.73 & $15.23^{+4.97}_{-5.10}$ & 1.4E-02$^\mathrm{+5.0E-03}_\mathrm{-5.1E-03}$ & $0.25^{+0.07}_{-0.06}$ & $7.44^{+0.41}_{-0.41}$ & $20.06^{+1.09}_{-1.01}$ & $176.09^{+0.96}_{-2.12}$ & $257.89^{+12.44}_{-13.13}$ \\
HD224538 & 359.72 & -61.59 & $7.03^{+1.34}_{-0.85}$ & 5.5E-03$^\mathrm{+1.3E-03}_\mathrm{-1.0E-03}$ & $0.48^{+0.01}_{-0.01}$ & $2.37^{+0.11}_{-0.12}$ & $3.29^{+0.01}_{-0.01}$ & $91.13^{+25.90}_{-25.50}$ & $214.93^{+52.01}_{-41.77}$ \\
HD23596 & 57.00 & 40.53 & $11.56^{+1.34}_{-1.42}$ & 1.0E-02$^\mathrm{+1.7E-03}_\mathrm{-1.8E-03}$ & $0.28^{+0.02}_{-0.01}$ & $2.69^{+0.11}_{-0.12}$ & $4.20^{+0.02}_{-0.03}$ & $38.90^{+15.76}_{-77.18}$ & $16.68^{+148.95}_{-21.64}$ \\
HD27894 & 65.20 & -59.41 & $6.78^{+0.70}_{-0.64}$ & 8.0E-03$^\mathrm{+1.3E-03}_\mathrm{-1.2E-03}$ & $0.34^{+0.03}_{-0.03}$ & $5.36^{+0.21}_{-0.22}$ & $13.74^{+0.09}_{-0.09}$ & $101.52^{+14.74}_{-31.50}$ & $129.18^{+4.14}_{-4.70}$ \\
HD29461 & 69.74 & 14.11 & $96.75^{+8.82}_{-8.72}$ & 8.9E-02$^\mathrm{+1.4E-02}_\mathrm{-1.4E-02}$ & $0.61^{+0.01}_{-0.01}$ & $4.90^{+0.19}_{-0.21}$ & $10.23^{+0.02}_{-0.02}$ & $88.65^{+18.68}_{-18.44}$ & $105.51^{+8.58}_{-8.89}$ \\
HD30177 & 70.48 & -58.02 & $8.69^{+0.96}_{-0.77}$ & 8.5E-03$^\mathrm{+1.4E-03}_\mathrm{-1.2E-03}$ & $0.21^{+0.01}_{-0.02}$ & $3.60^{+0.14}_{-0.15}$ & $6.88^{+0.01}_{-0.01}$ & $85.39^{+14.35}_{-18.74}$ & $311.23^{+65.06}_{-148.08}$ \\
HD30177 & 70.48 & -58.02 & $6.56^{+0.90}_{-0.75}$ & 6.4E-03$^\mathrm{+1.2E-03}_\mathrm{-1.0E-03}$ & $0.04^{+0.04}_{-0.03}$ & $10.27^{+0.52}_{-0.50}$ & $33.09^{+1.60}_{-1.21}$ & $98.02^{+16.02}_{-24.23}$ & $332.52^{+33.06}_{-23.90}$ \\
HD39091 & 84.30 & -80.47 & $12.24^{+1.28}_{-1.30}$ & 1.1E-02$^\mathrm{+1.8E-03}_\mathrm{-1.8E-03}$ & $0.64^{+0.00}_{-0.00}$ & $3.31^{+0.14}_{-0.15}$ & $5.72^{+0.00}_{-0.00}$ & $54.44^{+5.94}_{-3.72}$ & $278.55^{+6.68}_{-9.82}$ \\
HD62549 & 116.08 & -5.05 & $8.73^{+1.89}_{-1.25}$ & 7.7E-03$^\mathrm{+1.9E-03}_\mathrm{-1.5E-03}$ & $0.15^{+0.13}_{-0.10}$ & $11.79^{+2.20}_{-1.92}$ & $39.17^{+10.30}_{-8.82}$ & $162.31^{+3.61}_{-2.16}$ & $19.23^{+13.32}_{-11.88}$ \\
HD65430 & 119.89 & 20.84 & $105.25^{+8.52}_{-8.80}$ & 1.1E-01$^\mathrm{+1.7E-02}_\mathrm{-1.7E-02}$ & $0.38^{+0.00}_{-0.00}$ & $4.51^{+0.18}_{-0.19}$ & $9.64^{+0.00}_{-0.00}$ & $73.11^{+1.57}_{-1.05}$ & $8.11^{+0.87}_{-0.79}$ \\
HD6558 & 16.61 & -0.75 & $74.56^{+9.15}_{-8.71}$ & 6.4E-02$^\mathrm{+1.2E-02}_\mathrm{-1.1E-02}$ & $0.27^{+0.04}_{-0.04}$ & $6.62^{+0.32}_{-0.33}$ & $16.04^{+0.49}_{-0.45}$ & $7.08^{+0.39}_{-0.36}$ & $327.94^{+8.08}_{-7.42}$ \\
HD66428 & 120.87 & -1.16 & $10.78^{+2.61}_{-3.68}$ & 1.0E-02$^\mathrm{+2.8E-03}_\mathrm{-3.7E-03}$ & $0.47^{+0.01}_{-0.01}$ & $3.39^{+0.14}_{-0.15}$ & $6.21^{+0.01}_{-0.02}$ & $16.64^{+10.12}_{-2.96}$ & $299.39^{+8.49}_{-8.22}$ \\
HD73267 & 129.07 & -34.46 & $5.40^{+0.64}_{-0.55}$ & 5.5E-03$^\mathrm{+9.7E-04}_\mathrm{-9.1E-04}$ & $0.09^{+0.02}_{-0.02}$ & $12.62^{+0.70}_{-0.75}$ & $46.74^{+2.15}_{-2.98}$ & $91.85^{+18.59}_{-25.00}$ & $171.88^{+139.62}_{-128.41}$ \\
HD74014 & 130.33 & -4.81 & $61.52^{+5.63}_{-5.75}$ & 6.0E-02$^\mathrm{+9.7E-03}_\mathrm{-9.8E-03}$ & $0.53^{+0.00}_{-0.00}$ & $7.23^{+0.32}_{-0.34}$ & $19.21^{+0.30}_{-0.30}$ & $125.51^{+1.43}_{-1.41}$ & $158.31^{+6.05}_{-6.21}$ \\
HD74156 & 130.60 & 4.58 & $9.11^{+0.94}_{-0.91}$ & 7.9E-03$^\mathrm{+1.3E-03}_\mathrm{-1.3E-03}$ & $0.38^{+0.01}_{-0.01}$ & $3.68^{+0.15}_{-0.16}$ & $6.70^{+0.01}_{-0.01}$ & $120.16^{+7.60}_{-66.22}$ & $210.65^{+7.61}_{-42.04}$ \\
HD86264 & 149.24 & -15.90 & $12.86^{+8.70}_{-5.00}$ & 9.7E-03$^\mathrm{+6.7E-03}_\mathrm{-4.0E-03}$ & $0.82^{+0.10}_{-0.19}$ & $2.76^{+0.15}_{-0.15}$ & $4.08^{+0.16}_{-0.13}$ & $93.59^{+43.18}_{-50.05}$ & $102.05^{+82.73}_{-69.93}$ \\
HD8673 & 21.54 & 34.58 & $13.40^{+1.54}_{-1.57}$ & 1.0E-02$^\mathrm{+2.0E-03}_\mathrm{-2.0E-03}$ & $0.73^{+0.04}_{-0.03}$ & $2.97^{+0.15}_{-0.17}$ & $4.50^{+0.03}_{-0.04}$ & $95.45^{+19.44}_{-8.82}$ & $230.91^{+9.70}_{-40.68}$ \\
HD87883 & 152.18 & 34.24 & $5.37^{+0.51}_{-0.59}$ & 6.3E-03$^\mathrm{+9.3E-04}_\mathrm{-1.0E-03}$ & $0.71^{+0.01}_{-0.01}$ & $4.05^{+0.15}_{-0.16}$ & $9.06^{+0.08}_{-0.08}$ & $25.45^{+1.61}_{-1.05}$ & $99.95^{+4.23}_{-3.30}$ \\
HD88072 & 152.35 & 2.37 & $8.07^{+1.36}_{-1.18}$ & 7.4E-03$^\mathrm{+1.5E-03}_\mathrm{-1.4E-03}$ & $0.16^{+0.14}_{-0.10}$ & $13.91^{+4.12}_{-2.05}$ & $50.76^{+24.44}_{-10.18}$ & $169.63^{+1.39}_{-1.72}$ & $118.97^{+14.73}_{-15.07}$ \\
HD97037 & 167.55 & -7.39 & $21.00^{+2.07}_{-1.96}$ & 1.9E-02$^\mathrm{+3.0E-03}_\mathrm{-2.9E-03}$ & $0.36^{+0.03}_{-0.03}$ & $6.91^{+0.29}_{-0.31}$ & $17.66^{+0.39}_{-0.42}$ & $14.56^{+0.80}_{-0.71}$ & $359.43^{+3.48}_{-3.20}$ \\
HD98649 & 170.22 & -23.22 & $8.40^{+1.97}_{-1.43}$ & 7.9E-03$^\mathrm{+2.1E-03}_\mathrm{-1.6E-03}$ & $0.84^{+0.02}_{-0.02}$ & $6.29^{+0.34}_{-0.33}$ & $15.60^{+0.89}_{-0.79}$ & $78.84^{+52.24}_{-34.30}$ & $66.73^{+30.85}_{-36.35}$ \\
HIP22203 & 71.63 & 15.47 & $50.85^{+4.64}_{-4.63}$ & 4.7E-02$^\mathrm{+7.4E-03}_\mathrm{-7.4E-03}$ & $0.80^{+0.02}_{-0.02}$ & $2.01^{+0.08}_{-0.09}$ & $2.74^{+0.01}_{-0.00}$ & $84.36^{+8.68}_{-2.55}$ & $309.34^{+60.90}_{-81.69}$ \\
\hline
\end{longtable}

\begin{figure}
    \centering
    \includegraphics[width=0.5\textwidth]{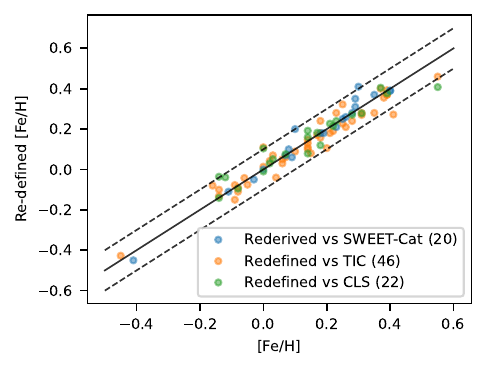}
    \caption{Comparison of our re-derived/redefined [Fe/H] with SWEET-Cat, TIC and CLS. The dashed line denotes the 0.1 dex difference.}
    \label{fig:stellar-feh}
\end{figure}

To homogeneously determine the metallicity of {54} host stars, we first make a cross match with SWEET-Cat database\cite{Sousa2021}\footnote{\url{http://www.astro.up.pt/resources/sweet-cat}}, whose spectroscopic stellar parameters were derived following the same homogeneous process using ARES \cite{Sousa2007A&A...469..783S,Sousa2015} and MOOG codes \cite{Sneden1973ApJ...184..839S}. 
A total of {28} planet-hosting stars are found listing in SWEET-Cat. Then we re-derive the metallicity of other {22} stars using the same codes. The high-resolution spectra of HARPS are obtained from the ESO archive data \footnote{Based on data obtained from the ESO Science Archive Facility with DOI(s): \url{https://doi.org/10.18727/archive/33}.}, and the ones with highest SNR are prioritized for processing. 
{The basics of spectral analysis is to realize the balance of iron abundance between excitation and ionization. The ARES\footnote{\url{https://github.com/sousasag/ARES}.} code can measure the equivalent widths of absorption lines, while the MOOG code determines element abundances using a grid of atmospheric model under the local thermodynamic equilibrium (LTE) assumption. Considering the potential difference in determining metallicity between ours and SWEET-Cat, we assign a conservative uncertainty of 0.1 dex to our re-derived values.}
For the rest of {4} stars, we carefully choose their metallicity values from the SIMBAD database \cite{wenger00} that derived with high-resolution spectra. To further validate our method, we additionally selected 20 host stars from our sample that are listed in SWEET-Cat and have available HARPS spectra, and re-derived their metallicity. As shown in Fig.~\ref{fig:stellar-feh}, our results show great consistency with SWEET-Cat, and the difference between our re-defined metallicity and those from TIC and the California Legacy Survey \citep[CLS;][]{rosenthalCaliforniaLegacySurvey2021} is not more than $1\sigma$. 

\begin{figure}
    \centering
    \includegraphics[width=0.6\textwidth]{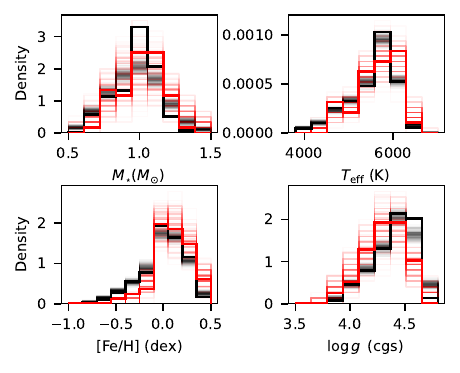}
    \caption{Individual histograms for stellar mass, effective temperature, metallicity and surface gravity of our sample. The total selected sample and the sample with detected companions are indicated by thick black and red lines, respectively. The light-colored lines in each panel are 200 histograms randomly drawn from normal distributions by accounting for the individual measurement uncertainties.}
    \label{fig:stellar-par}
\end{figure}

\begin{figure}
    \centering
    \includegraphics[width=0.5\linewidth]{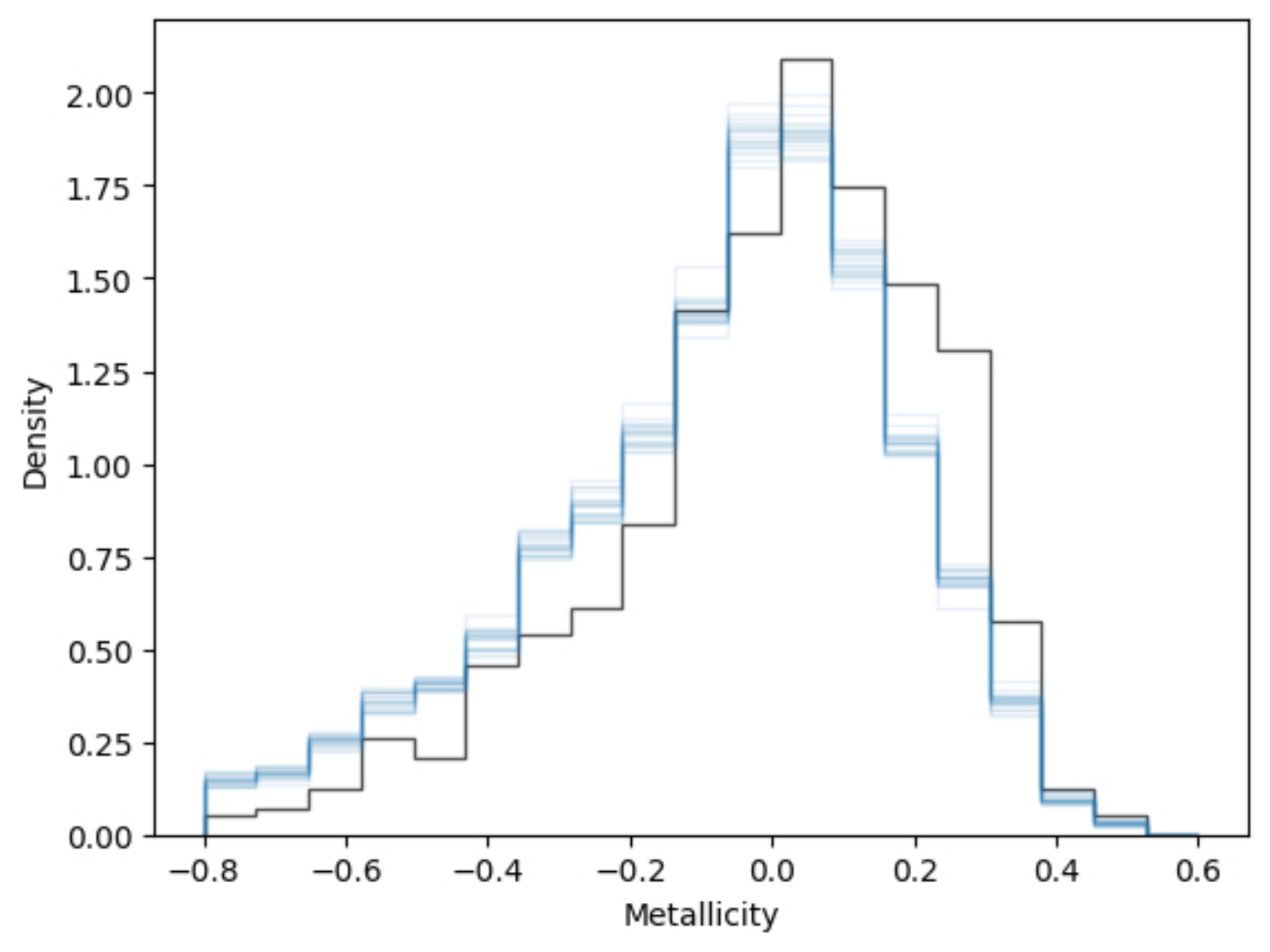}
    \caption{The metallicity distributions of resampled Gaia 100\,pc sample (blue histogram with 20 random draws) and our stellar sample (black histogram).}
    \label{fig:resample-gaia}
\end{figure}

Fig.~\ref{fig:stellar-par} shows the histograms for stellar mass, effective temperature, metallicity and surface gravity. 
These stars have a median temperature of 5665\,K and a median mass of $1.01\,M_{\odot}$ that are close to the solar properties. Their metallicity closely matches that of the unbiased 100\,pc FGK sample from Gaia (see Fig.~\ref{fig:resample-gaia}), further supporting that our RV targets are likely free from selection bias.

\section*{Detection efficiency and reliability}

To assess the detection efficiency for each star in the sample, we employ the injection-recovery technique. This involves synthesizing planet signals across various planetary masses and orbital elements to evaluate the likelihood of detection.

\subsection*{Injection of signals}
{For 54 systems with identified companions,} we initially subtract the planet's signal from the RV time series using the optimal orbital solution to obtain residual data. These residuals can be {approximately} regarded as pure noise caused by the star itself or instruments. 
Next, we inject synthesized planetary signals into these residuals {or RVs of other 736 systems that do not have known planetary signals,} using the following formula  
\begin{equation}
    {\rm RV}_{j} = K\,[{\rm cos}(\omega+\nu_{j})+e\,{\rm cos}(\omega)],
    \label{formula:RVs}
\end{equation}
where $K$ is the semi-amplitude of the Keplerian radial velocity variation, $\omega$ is the argument of periastron of the stellar reflex motion, $\nu_{j}$ is the true anomaly at each epoch $t_j$, and $e$ is the eccentricity. The true anomaly $\nu_{j}$ is related to the eccentric anomaly $E_{j}$, and the latter is related to the mean anomaly $M_j$ according to Kepler’s equation. Their relations can be expressed as:
\begin{equation}
    M_j=\frac{2\pi}{P}(t_j-T_{p}),
\end{equation}
\begin{equation}
    M_j=E_j-e\,{\rm sin}\,E_j,
\end{equation}
\begin{equation}
    {\rm tan}\frac{\nu_j}{2}=\sqrt{\frac{1+e}{1-e}}\cdot {\rm tan}\frac{E_j}{2},
\end{equation}
where $T_{p}$ is the epoch of periastron passage, $P$ is the orbital period. The semi-amplitude can be written as 
\begin{equation}
    K\equiv\frac{m\,{\rm sin}\,I}{(M_{\star}+m)^{\frac{1}{2}}}\left[\frac{G}{a(1-e^2)}\right]^{\frac{1}{2}},
    \label{formula:Kamp}
\end{equation}
where $a$ is the semi-major axis of the secondary relative to the primary star, $I$ is the inclination, and $m$ and $M_{\star}$ are respectively the masses of the companion and the star.
The $m$ and $a$ are uniformly sampled in logarithmic space {($50\times50$ bins)}, while other orbital parameters, aside from $T_{p}$ and $M_{\star}$ that fix to specific values of each star, are uniformly sampled {($50\times50\times30$)}, as shown in Table \ref{tab:samples}. {The $m$ and $a$ can be easily converted to mass ration $q$ and orbital period $P$ based on Kepler's third Law.}
Following ref.\cite{feng16}, we calculate the logarithmic Bayes Factor (ln\,BF) for a given dataset to assess the significance of an injected signal. Those with ln\,BF large than 5 will be selected for subsequent recovery analyses.

\begin{table*}
     \centering
     \caption{Sampling parameters and ranges}
     \begin{tabular*}{\textwidth}{@{}@{\extracolsep{\fill}}lcl@{}}
         \hline
         Parameter       & Distribution  & 
         Description     \\\hline
         $m$ ($M_{\rm J}$)    & Log-$\mathcal{U}^{[1]}$(3, 200)  & Companion mass  \\
         $a$ ($\rm au$) & Log-$\mathcal{U}$(1.2, 45) & Semi-major axis\\
         $I$ (radian)   & Cos-$\mathcal{U}^{[2]}$(-1, 1) & Inclination  \\        
         $e$    & $\mathcal{U}^{[3]}$(0, 1)  & Eccentricity  \\
         $\omega$ (radian)    & $\mathcal{U}$(0, 2$\pi$) &Argument of periapsis   \\
         $\Omega$ (radian)   & $\mathcal{U}$(0, 2$\pi$) &Longitude of ascending node   \\
         \hline
     \end{tabular*}
     \begin{tablenotes}
     \item[1] [1] Log-$\mathcal{U}$($x_1$\ , \ $x_2$) stands for a log-uniform distribution between $x_1$ and $x_2$.
     \item[2] [2] Cos-$\mathcal{U}$($x_1$\ , \ $x_2$) stands for a uniform distribution between $x_1$ and $x_2$ in cosine space.
     \item[3] [3] $\mathcal{U}$($x_1$\ , \ $x_2$) stands for a uniform distribution between $x_1$ and $x_2$.
     \end{tablenotes}
     \label{tab:samples}
 \end{table*}

We also simulate astrometric signals for each star simultaneously.
We utilize epoch and scan law provided by Hipparcos Intermediate Astrometry Data (IAD\cite{vanLeeuwen2007}) and Gaia Observation Forecast Tool \footnote{\url{https://gaia.esac.esa.int/gost/index.jsp}} (GOST) to sample the stellar projected orbits on the celestial sphere, thus obtaining synthesised astrometric data. Hipparcos IAD and GOST data comprise observation time, scan angle and parallax factor and other auxiliary information for a specific star, that will be incorporated into our injection modelling. 

In the plane of the sky, the projected offsets of a star's reflex motion relative to the system's barycenter are given by
\begin{equation}
    \Delta\alpha \ast=(\frac{m}{M_{\star}+m})(BX+GY)
    \label{formula:alpha}
\end{equation}
\begin{equation}
    \Delta\delta=(\frac{m}{M_{\star}+m})(AX+FY).
    \label{formula:delta}
\end{equation}
where $\Delta\alpha \ast$ and $\Delta\delta$ are respectively the offset in right ascension and declination, $A$, $B$, $F$ and $G$ are the Thiele-Innes coefficients. For dark companions, their luminosity can be omitted, {so that} the photocenter motion is equivalent to reflex motion.
The elliptical rectangular coordinates $X$ and $Y$ are functions of the eccentric anomaly $E_{j}$ and eccentricity $e$, given by
\begin{equation}
    X={\rm cos}\,E_{j}-e~,
\end{equation}
\begin{equation}
    Y=\sqrt{1-e^2}\cdot {\rm sin}\,E_{j}.
\end{equation}

Then we synthesize the along-scan (AL) position of a star (i.e., abscissa; $\eta_j$) by projecting the stellar offsets onto the 1D AL direction using 
\begin{equation}
    \eta_{j}=\Delta\alpha\ast{\rm sin\,}\psi+\Delta\delta {\rm cos\,}\psi,
\end{equation}
where $\psi$ is the scan angle of Gaia at epoch $t_j$. This scan angle is the complementary angle for Hipparcos, and thus will be changed to ($\pi/2-\psi$) when synthesizing Hipparcos abscissa. Because the parallax and proper motion of the barycenter are generally independent of the reflex motion, the parallax of a star is fixed to the catalog value of Gaia DR3, and the linear motion of the system's barycenter is ignored to simplify the injection and recovery process. 
Therefore, the parallax term and the proper motion of the barycenter in the above formula can be omitted, which means only stellar reflex motion is considered.

\subsection*{Recovery of signals}
After injecting signals into both radial velocity and astrometry data, we can recover the planetary signals using a simplified algorithm based on our detection pipeline. First, we detect radial velocity signals with a Keplerian model (see eq. \ref{formula:RVs}) with ${\rm ln\,BF}>5$ as a threshold to determine if a signal is significant\footnote{Using the python \texttt{scipy.optimize.curve\_fit} library}. The best-fit value of $K$, $P$, $e$, $\omega$ and their corresponding errors are obtained accordingly. For signals with orbital period longer than the observational baseline, we additionally require points of inflection existing in the synthetic RVs. Otherwise we perform a second order polynomial fitting to exclude those with only linear trend (SNR$<$3). The SNR are defined as $a_2/\Delta a_2$, where $a_2$ and $\Delta a_2$ respectively represent the second order coefficient and its associated error. 

Once we find a significant radial velocity signal, we start to fit synthesised Hipparcos and Gaia astrometric abscissae separately using a four-parameter linear model to obtain positions ($\hat{\alpha} $, $\hat{\delta}$) and proper motions ($\hat{\mu}_{\alpha}$, $\hat{\mu}_{\delta}$) of a star at reference epochs ($t_{\rm Hip}$, $t_{\rm DR3}$). 
The model is expressed as
\begin{equation}
    \hat{\eta}_{j}=(\hat{\alpha}+\hat{\mu}_{\alpha})\,{\rm sin\,}\psi+(\hat{\delta}+\hat{\mu}_{\delta})\, {\rm cos\,}\psi.
\end{equation}
Then we can obtain two sets of model parameters for Hipparcos ($\hat{\alpha}_{\rm Hip}$, $\hat{\delta}_{\rm Hip}$, $\hat{\mu}_{\alpha \rm Hip}$, $\hat{\mu}_{\delta \rm Hip}$) and Gaia ($\hat{\alpha}_{\rm DR3}$, $\hat{\delta}_{\rm DR3}$, $\hat{\mu}_{\alpha \rm DR3}$, $\hat{\mu}_{\delta \rm DR3}$), respectively, and their corresponding uncertainties are adopted from catalogue astrometry. We focus on the differences in star positions and proper motions between these two astrometry. Assuming that these positions and proper motions represent the instantaneous status of the stellar reflex motion at two reference epoch, we can analytically derive or recover the Thiele-Innes coefficients $A$, $B$, $F$ and $G$ through formulae (\ref{formula:alpha}) and (\ref{formula:delta}), and their errors can be obtained through the error propagation formulae. These coefficients can then be converted into Campbell elements using the \texttt{nsstools} code\footnote{\url{https://gitlab.obspm.fr/gaia/nsstools}}\cite{Halbwachs2023}, which allow us to calculate the orbital inclination $I$ along with its error. The basic formula is 
\begin{equation}
    I={\rm arccos}\left(\frac{c_1}{c_2+\sqrt{c_2^2-c_1^2}}\right),
\end{equation}
where $c_1=AG-BF$ and $c_2=(A^2+B^2+F^2+G^2)/2$. Finally, we combine inclination $I$ with semi-amplitude $K$ from radial velocity fitting to estimate planet mass error ($\Delta m$) through formula (\ref{formula:Kamp}). The final recovered planetary signal requires that the relative mass error ($\Delta m/m$) of planets should be less than {60\%}. 
This threshold value is applied in our original selections of companion sample with reliable orbital solution\cite{feng3DSelection1672022}.

\subsection*{Detection efficiency}

\begin{figure}
    \centering
    \includegraphics[width=0.5\textwidth]{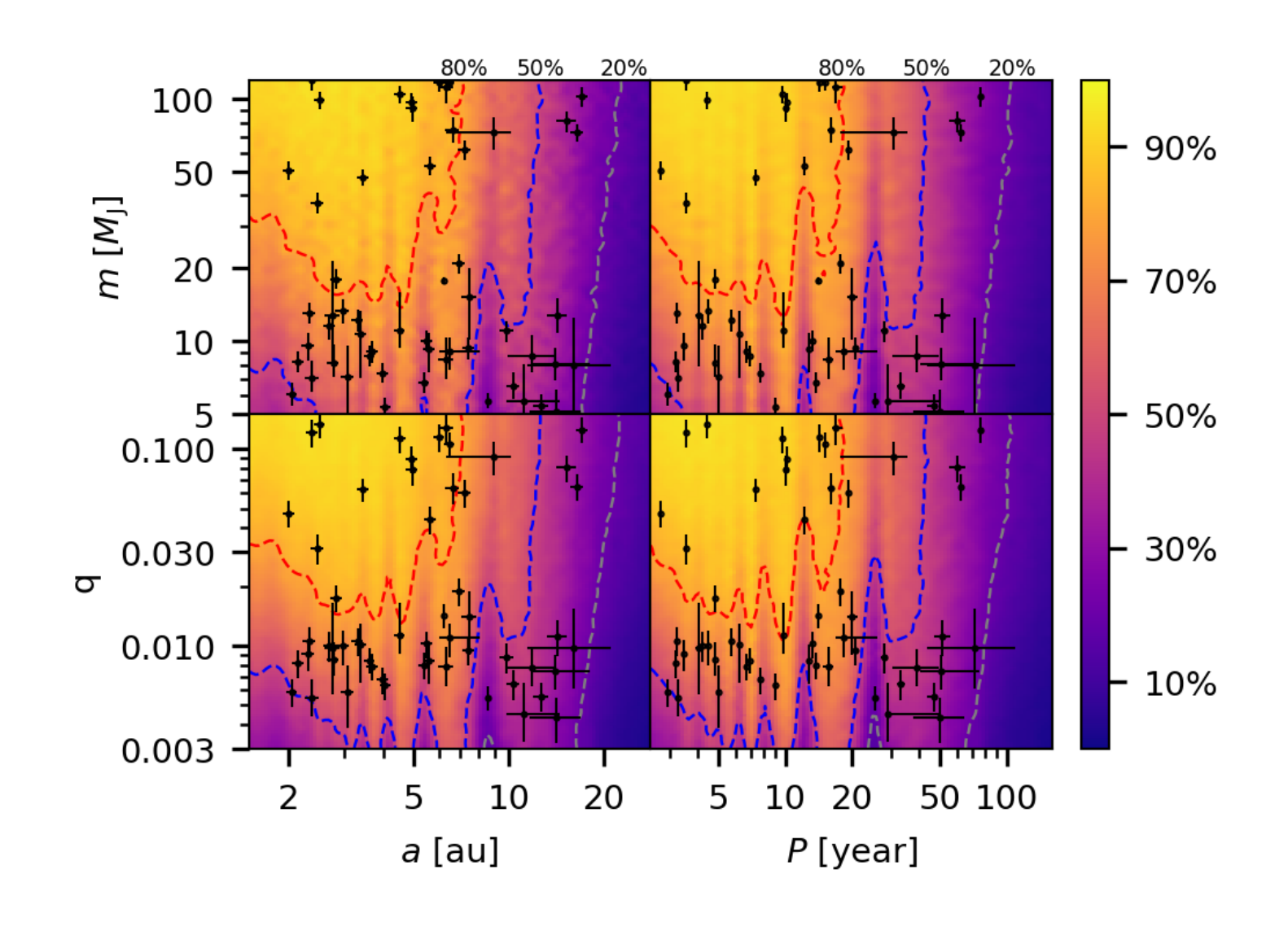}
    \caption{Distribution of samples alongside detection efficiency across various parameters: companion mass $m$, semi-major axis $a$, mass ratio $q$, and orbital period $P$. Observed data points are denoted by scattered black markers with error bars, while the detection efficiencies are illustrated by colour gradients in the background.}
    \label{fig:completeness}
\end{figure}

For each star, we inject {75,000} trial signals and repeat above recovery procedure. We make a grid of bins in our 4-dimension parameter space ($a\text{--}P\text{--}m\text{--}q$), and count the numbers of recovers divided by the injected sources as the detection efficiency. The individual detection efficiency can be combined to illustrate the overall detection capabilities of the survey.
In the result, Fig.~\ref{fig:completeness} shows the detection efficiency in the $m\text{--}a$, $m\text{--}P$, $q\text{--}a$, and $q\text{--}P$ space, respectively. The right regions with low probability in each panel indicate that more planets have remained elusive due to insufficient radial velocity observations, while the left regions are subject to the precision of Hipparcos-Gaia astrometry. Meanwhile, a set of significant harmonic-like regions is seen in these efficiency maps, which can be attribute to the $\sim$25 years temporal baseline between Hipparcos and Gaia DR3. When an injected period is an integer fraction of the baseline, Hipparcos and Gaia satellites will sample an arc with nearly identical orbital phases. This makes it difficult to constrain the orbital inclinations, thereby decreasing the detection efficiency of these regions.

\subsection*{Detection reliability}
\begin{figure}[h]
    \centering
    \includegraphics[width=0.5\textwidth]{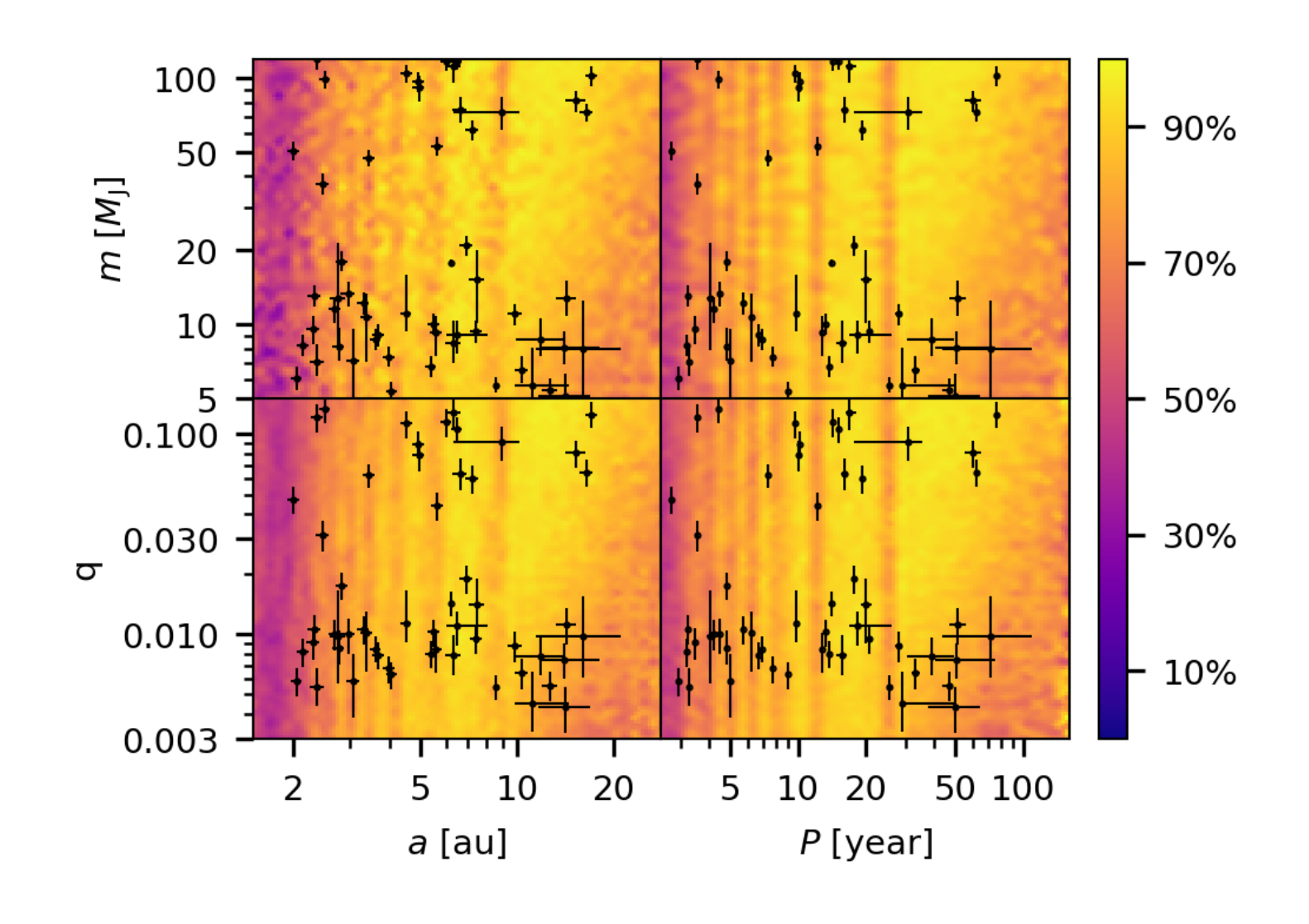}
    \caption{Distribution of samples alongside detection reliability across various parameters. The symbols are similar to Extended Fig.\ref{fig:completeness}}
    \label{fig:reliability}
\end{figure}

We also evaluate the reliability of our detection method. {Unlike the efficiency analysis, which covers all 790 stars, we restrict the reliability study to the 54 systems that host companions.}
The reliability is calculated by dividing the number of valid signals by the total recovered signals in each bin. We additionally require that the valid signals should satisfy: 1) the relative differences for semi-amplitude $K$, period $P$ {and companion mass $m$} between injected and recovered signal are less than 30\%, 2) the eccentricity differences are within 0.1, and 3) the inclination differences are within $30\deg$. As presented in Extended Fig.~\ref{fig:reliability}, almost all the quantities across various parameters are larger than 50\%, suggesting the advantages and reliability of our detection method. 

\subsection*{Potential biases}
\begin{figure}
    \centering
    \includegraphics[width=0.5\textwidth]{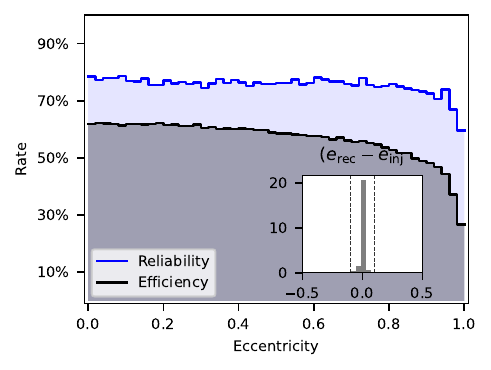}
    \caption{Detection efficiency and reliability for eccentricity. The bottom histogram shows the discrepancy between recovered and injected values of eccentricity. Two dashed lines indicate the threshold ($0.1$) of reliable detection.}
    \label{fig:detection-rate-e}
\end{figure}

\begin{figure}[h]
    \centering
    \includegraphics[width=0.8\textwidth]{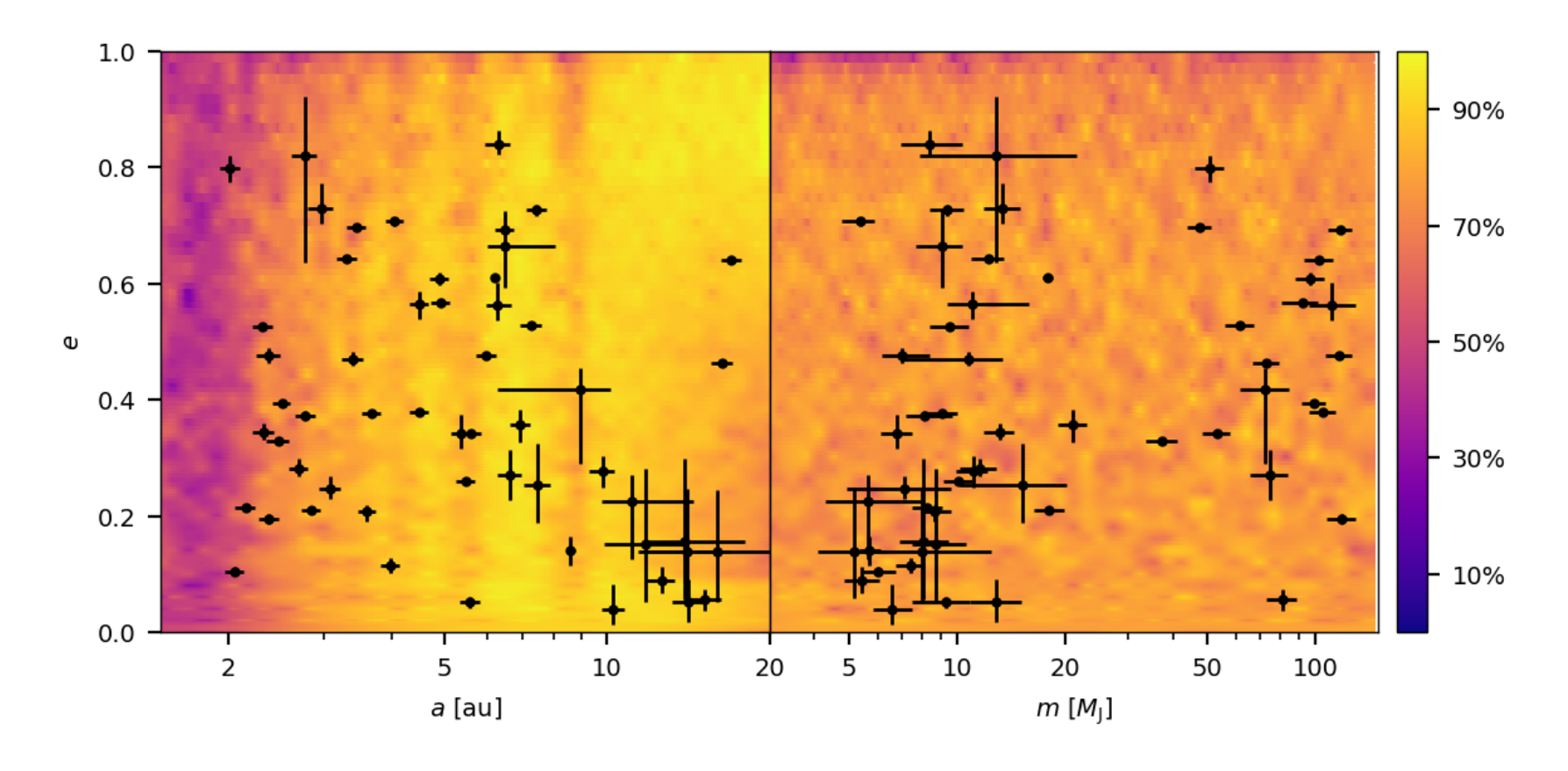}
    \caption{Distribution of samples alongside detection reliability across $a-e$ and $m-e$ parameter spaces. The symbols are similar to Extended Fig.\ref{fig:completeness}}
    \label{fig:a-m-e}
\end{figure}

Additionally, we also investigate the potential biases of our method in determining eccentricity and inclination.
As shown in Fig.~\ref{fig:detection-rate-e}, the detection efficiency for signals with $e<0.8$ is close to a uniform distribution. It falls rapidly as $e$ increases {beyond 0.8}, suggesting that our method might be not sensitive to {highly eccentric} planets. This can be anticipated as the orbital phase of long-period, high-eccentricity companions {is likely} near the apastron, and therefore can be easily misinterpreted as relatively low $e$ {companions}.  
We however note that the detection reliability in $e$ space is as high as {$\sim$80\%} across the full range of  eccentricities, and 98\% of recovered signals have a discrepancy not exceeding $0.1$ with injected $e$. 
Therefore, it is very likely that the observed low-$e$ population in Fig.~\ref{fig:ecc} {reveals} the real nature of wide-orbit planets instead of {caused by} detection biases. 
{To corroborate this further, we present the reliability map in the $a-e$ and $m-e$ planes in Fig.~\ref{fig:a-m-e}. Evidently, the reliability for long-period, low-mass, and low-$e$ companions exceeds $60\%$.}

\begin{figure}
    \centering
    \includegraphics[width=0.5\textwidth]{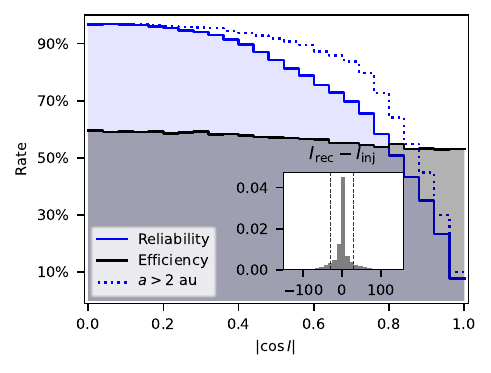}
    \caption{Detection efficiency and reliability for $|\cos I|$. The dotted line denotes the reliability for systems with $a>2\,au$. The bottom histogram shows the discrepancy between recovered and injected values of inclination. Two dashed lines indicate the $30\deg$ threshold of reliable detection. We here utilise the absolute value of $\cos I$ to take into account the bimodality of our measured inclinations.}
    \label{fig:detection-rate-cosi}
\end{figure}

When considering inclinations, as shown in the Fig.~\ref{fig:detection-rate-cosi}, we find the detection efficiency for edge-on systems ($|{\rm cos}\,I|\sim0$) is slightly higher than face-on systems ($|{\rm cos}\,I|\sim1.0$). This is primarily due to the fact that the radial velocity variations caused by a planet in a face-on orientation have diminished. Similar trend can be found for reliability, but it is steeper towards to the relative face-on systems. Likewise, {91\%} of recovered inclinations align well with the injected one within $30\deg$. We further find that the relatively lower reliability of face-on systems mainly occurs in regions with $a<2\,{\rm au}$ (Fig.~\ref{fig:reliability}), which suffers from the $\sim$1034\,days baseline of Gaia DR3. Beyond this separation, the reliability becomes less steep.

\begin{figure}
    \centering
    \includegraphics[width=0.5\textwidth]{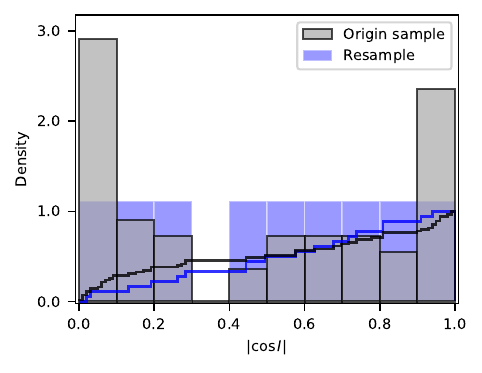}
    \caption{Distribution of $|{\rm cos}\,I|$. Our original sample and the resample of $|{\rm cos}\,I|$ are respectively coloured in grey and blue, and the thick lines denote their CDF. The uncertainties of $I$ for each system are considered.}
    \label{fig:distribution_inc}
\end{figure}

{The above analyses suggest that there is no strong evidence for significant biases in our method of determining inclinations.}
Therefore, it is expected that our observed inclinations should be nearly random in distribution. However, as stated in ref.\cite{benedict14HerPlanetary2023}, both our previous work of ref.\cite{feng3DSelection1672022} and HST-derived inclinations are not as random as the binary star (6th Catalog of Visual Binary Stars\cite{hartkopf2001USNaval2001}) inclinations due to some unresolved issues. Those systems selected for HST astrometry usually have a prior knowledge of substellar companions from precision radial velocity investigations\cite{benedictAstrometryHubbleSpace2017}.  
Extended Fig.~\ref{fig:distribution_inc} shows the distribution of our $|{\rm cos}\,I|$, in which we fold all inclinations over $90\deg$ into the $0\sim90\deg$ scale (i.e., $180\deg-i$) since a fraction of our companions show bimodality in inclination. This bimodality has no effect on the determination of companion mass as it mainly relates to $\sin I$.
It {appears} that our inclinations apparently deviate from an isotropic distribution and exhibit some excess for face-on systems, which differs from HST-derived inclinations of skewing to small values\cite{benedictAraePlanetarySystem2022}. We found a $p$-value of {0.029} from a Kolmogorov-Smirnov test\footnote{Using the python \texttt{scipy.stats.kstest} library} that rejects the null hypothesis that our inclinations are drawn from an isotropic distribution. 
Multiple factors, including limited statistical data, unresolved aspects of our analyses, or inherent characteristics of our solar neighbourhood, could contribute to this {discrepancy}. 
However, deciphering the truth necessitates additional observational data and a more extensive sample of planets with accurately determined inclinations.

\begin{figure}
    \centering
    \includegraphics[width=0.5\textwidth]{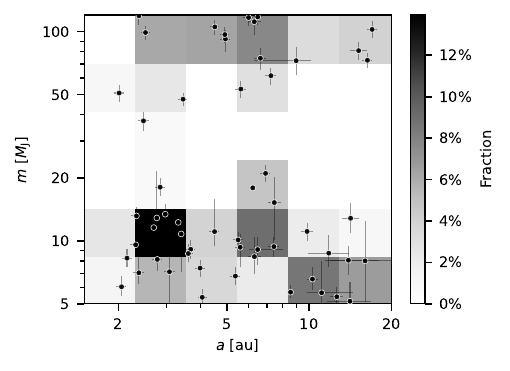}
    \caption{2D histogram in $m\text{--}a$ plane. The observed data points are denoted by scattered black markers with error bars, while the fractions after 100 resampling are illustrated by colour gradients in the background.}
    \label{fig:depandent_inc}
\end{figure}

To investigate potential dependencies between the significant features identified in Fig.~\ref{fig:weighted-KDE} and the heterogeneous distribution of inclinations, we initially partition $|{\rm cos}\,I|$ into 10 bins based on the median error of $\Delta{\rm cos}\,I\sim0.1$. From each bin, we randomly select 2 systems to construct a relatively isotropic subsample. 
Subsequently, we conduct 100 resamplings and project all selected systems onto the $m\text{--}a$ plane. Their cumulative distribution function (CDF) is illustrated in Extended Fig.~\ref{fig:distribution_inc} and the 2D histogram is plotted as the background of Extended Fig.~\ref{fig:depandent_inc}. {It is clear that the brown dwarf desert is still significant and thus show no relevance with inclinations. }
The quantitative analyses are presented in subsequent sections.

{Because we inject only one companion at a time, our injection-recovery test does not account for multiplicity. This may introduce some potential bias. However, since the companions we detect usually have much longer periods than inner companions, this bias is likely to be insignificant. Nonetheless, we acknowledge that, in reality, mutual interactions, especially in systems where companions are more closely spaced or possess near integer orbital period ratios, may introduce additional complexities not captured by our current framework.}

\section*{Occurrence rate results}
\begin{figure}[!h]
    \centering
    \includegraphics[width=0.7\linewidth]{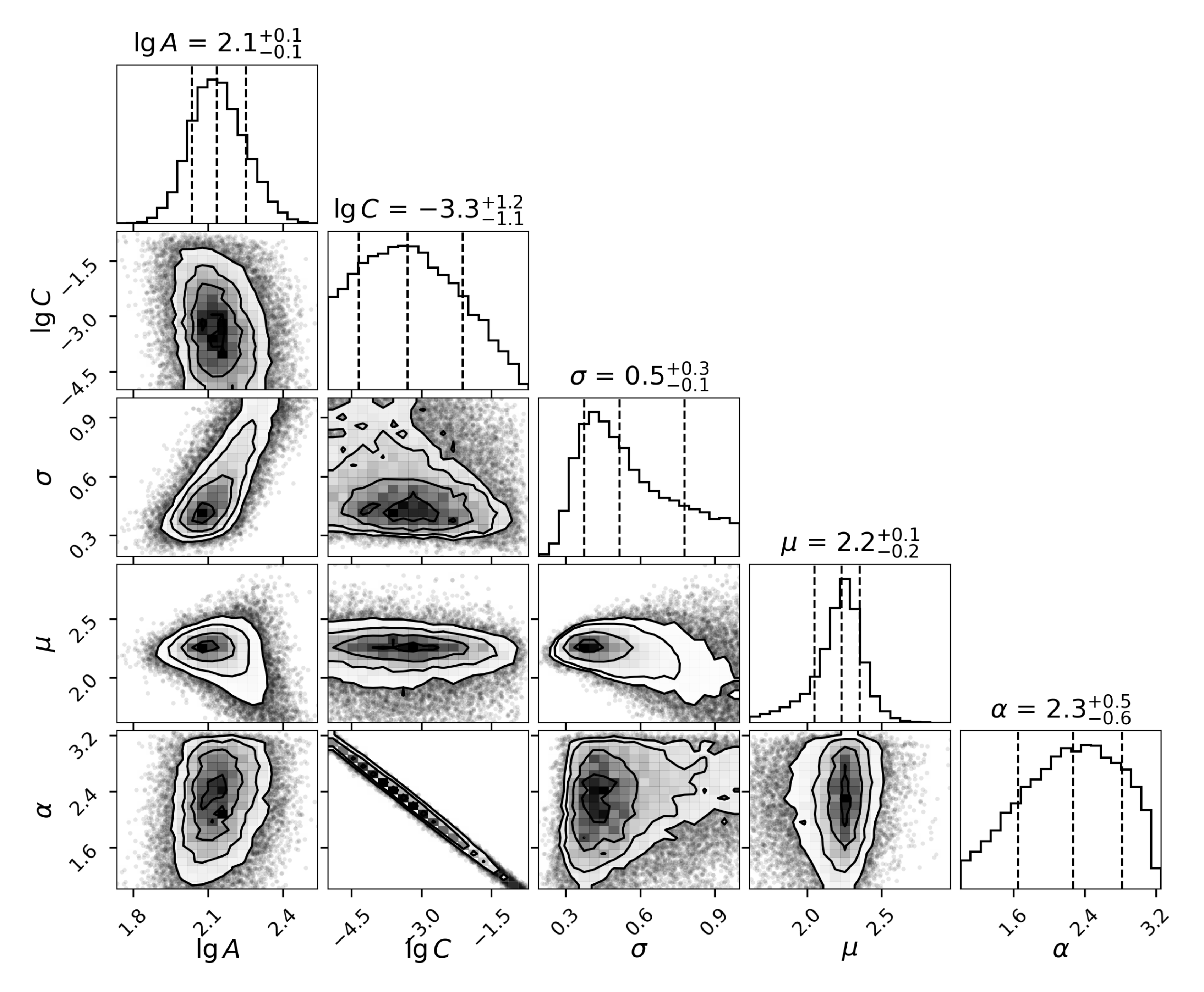}
    \caption{Marginalised posterior probability distributions of parameters of 1D parametric model for power law and log-normal.}
    \label{fig:1d-le17au-parametric}
\end{figure}

Fig.~\ref{fig:1d-le17au-parametric} shows the corner plot of the MCMC parameters' marginal distributions for our 1D log-normal and power law model.

\begin{figure}
    \centering
    \includegraphics[width=1\linewidth]{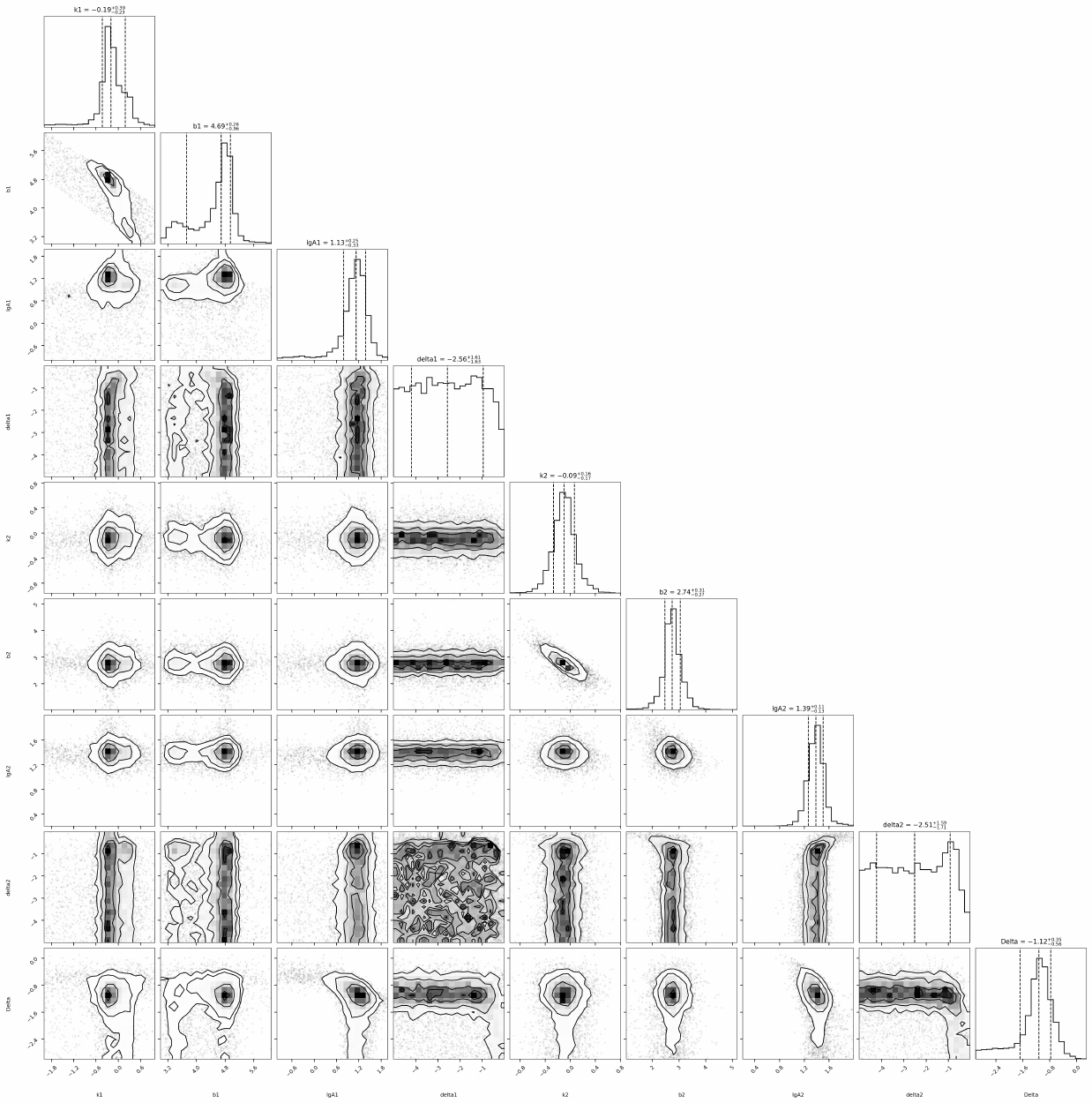}
    \caption{{Marginalized posterior probability distributions of parameters of the boundary model.}}
    \label{fig:boundary-corner-plot}
\end{figure}

\section*{Metallicity and eccentricity distributions for different groups}

\begin{figure}
    \centering
    \includegraphics[width=0.5\linewidth]{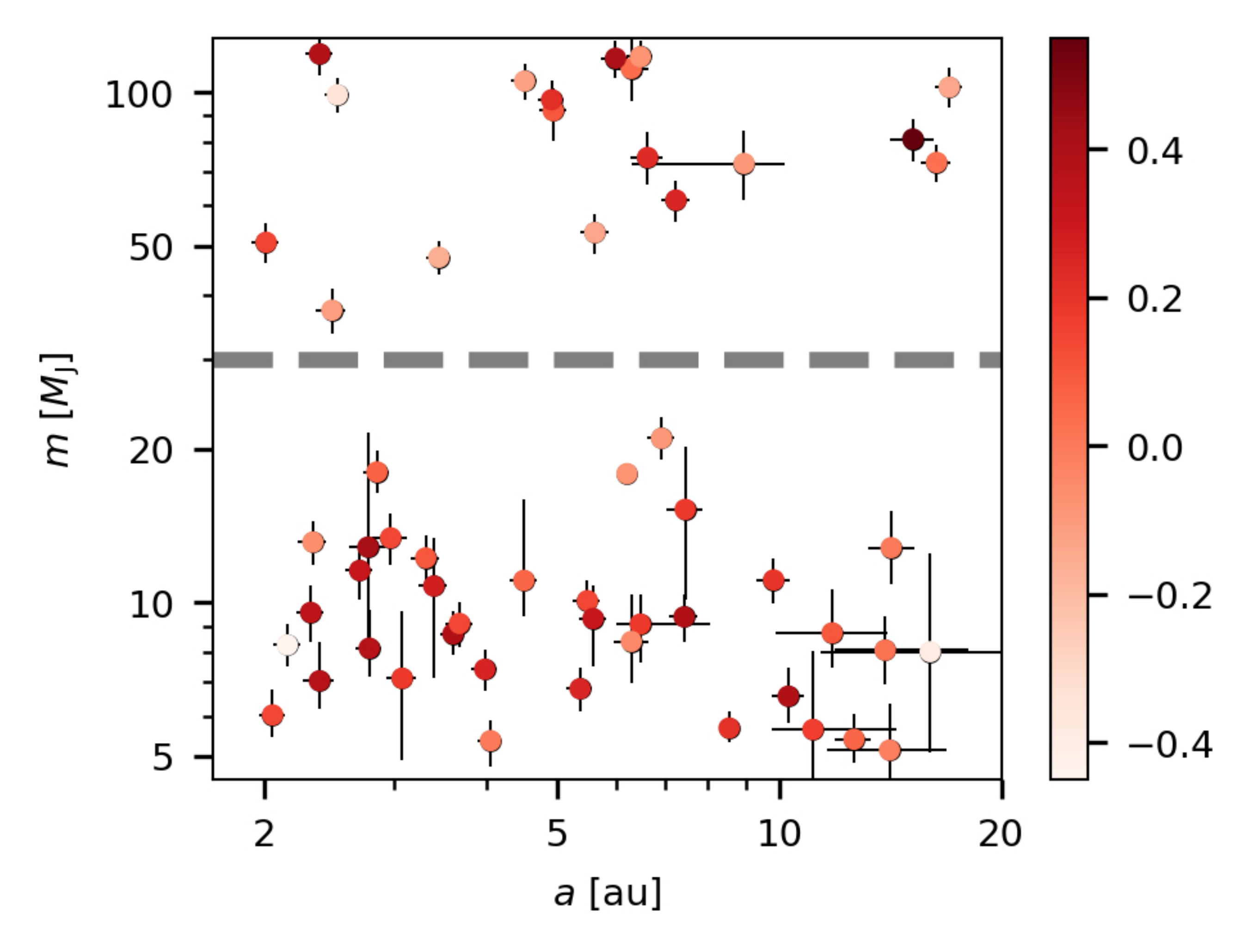}
    \caption{Metallicity distribution of the selected sample. Similar to Extended Fig. \ref{fig:ecc}, but color-mapped with metallicity.}
    \label{fig:feh}
\end{figure}

The metallicity distributions of three groups are smoothed with KDE. The KDE bandwidth equals to 0.25. The known planetary hosts from exoplanet archive data are obtained at January 9th, 2024. We use the Kolmogorov-Smirnov test to compare the metallicity of group A with that of other groups and all known planet hosts, operating under the null hypothesis that group A has greater CDF and alternative hypothesis that group A has lower CDF. As a reference, we show our metallicity distribution in the $m\text{--}a$ plane (the Extended Fig.~\ref{fig:feh}).

\begin{figure}
    \centering
    \includegraphics[width=0.5\linewidth]{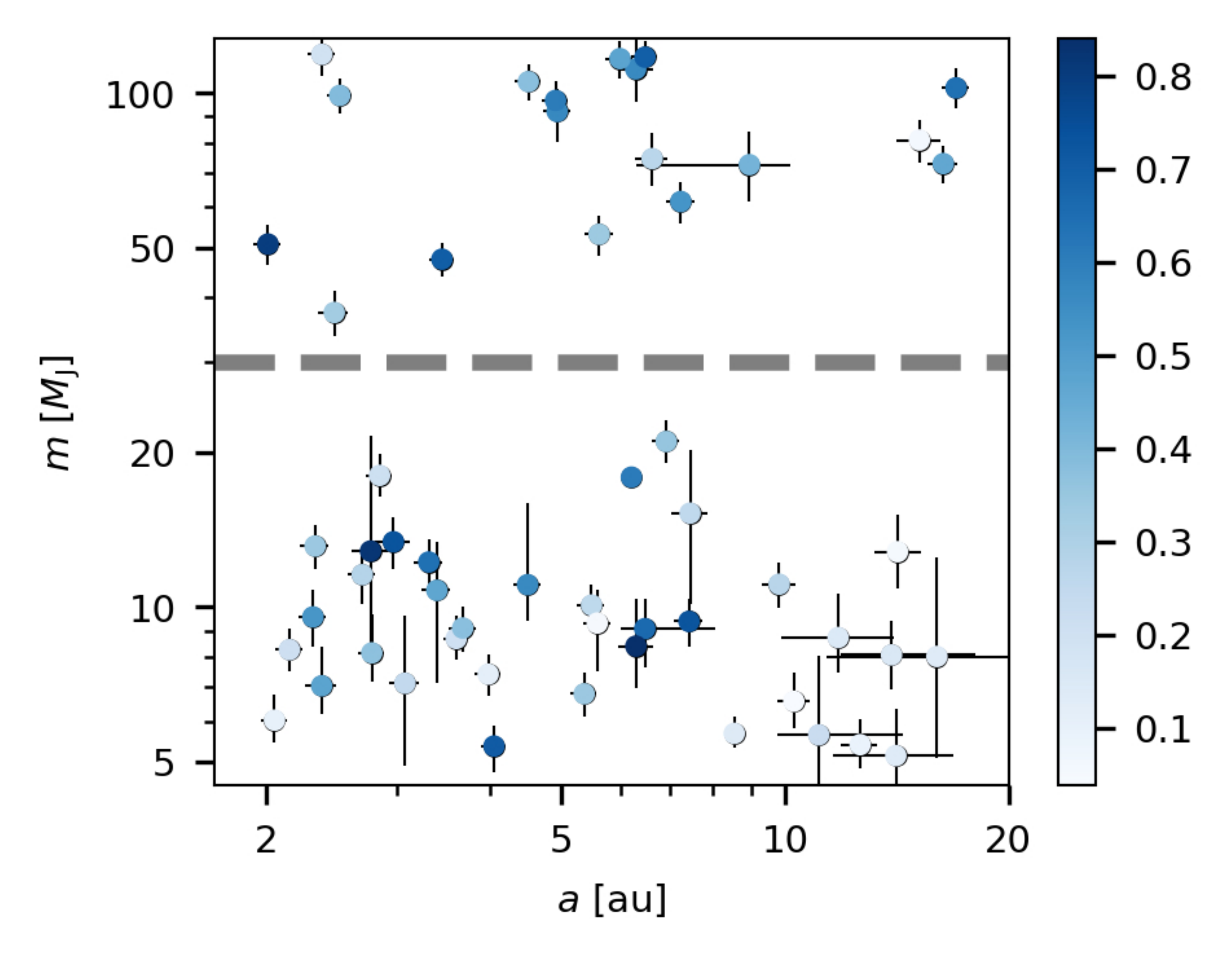}
    \caption{Eccentricity color-coded of the selected sample. The dashed gray line indicates the 30 $\mj$.}
    \label{fig:ecc}
\end{figure}

The eccentricity and metallicity distributions for two mass ranges are shown in Fig. \ref{fig:feh-ecc-dist}. These distributions were calculated using the inverse detection rate, employing methodologies consistent with detection efficiency-weighted KDE.
As a reference, we also show our eccentricity and metallicity distributions in the $m\text{--}a$ plane (the Fig.~\ref{fig:ecc} and \ref{fig:feh}).

\begin{figure}
    \centering
    \includegraphics[width=0.6\linewidth]{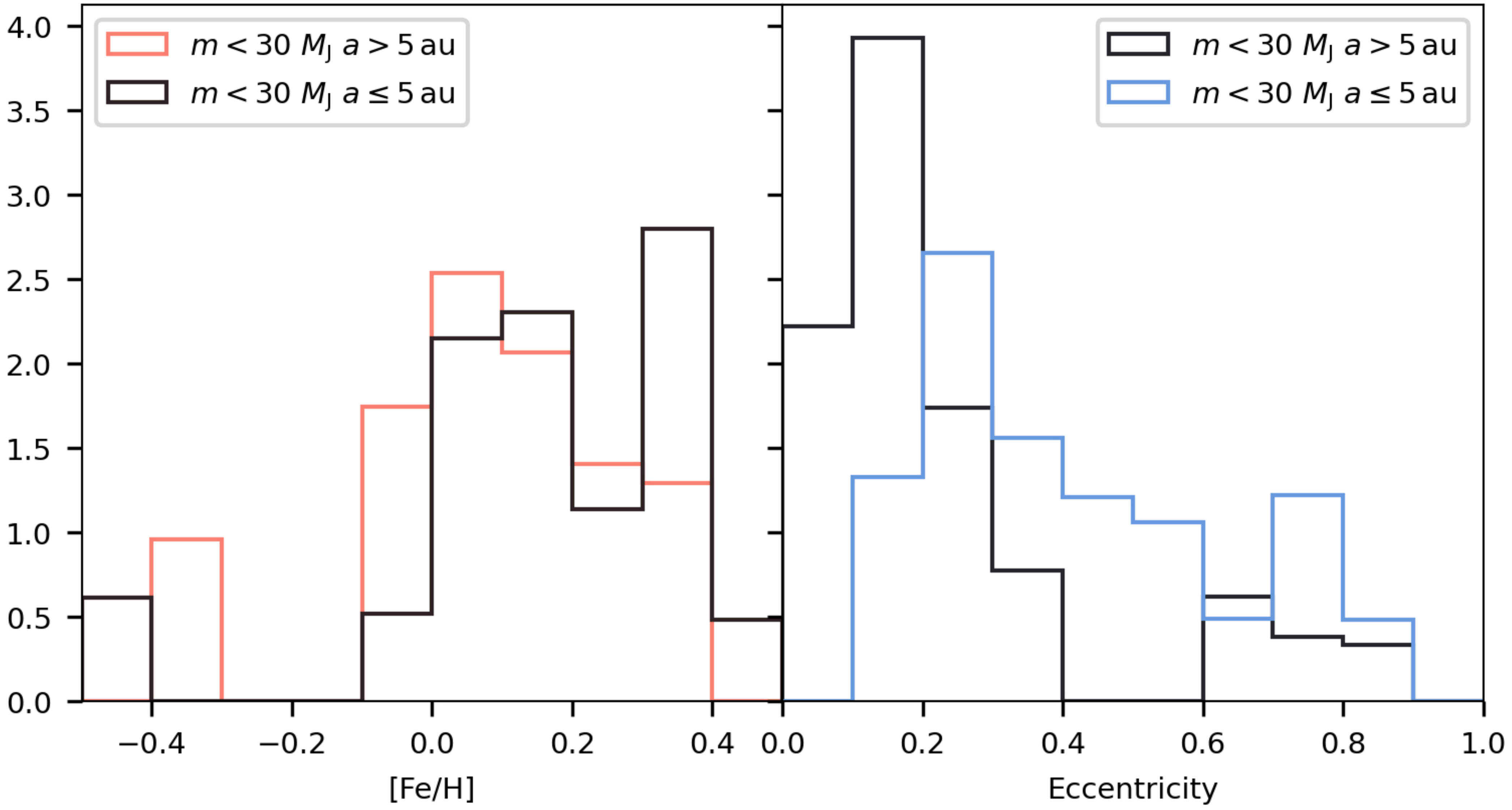}
    \caption{Histograms of metallicities and eccentricities for the inner ($\leq 5\,$au) and outer ($> 5\,$au) low mass companions ($m < 30\,\mj$).}
    \label{fig:hist-ecc-feh-le30mj}
\end{figure}

Furthermore, there is a weak differences inside the lower mass companions ($m < 30\,\mj$). We split them by 5\,au. One can see their distributions of metallicity and eccentricity (Fig.~\ref{fig:hist-ecc-feh-le30mj}).

\section*{Population synthesis}
Our population synthesis study incorporates two models: CA and GI. The CA model, despite being the most conventional model for planet formation, primarily yields giant planets but struggles to replicate our observation  when used in isolation.

\subsection*{Core accretion model} 
We employ a Monte Carlo sampling approach to conduct the pebble-driven core accretion planet population synthesis model. This involves sampling the initial disc, the planet embryos, and the stellar properties of systems. 
For instance, the host stellar mass and metallicity are randomly selected from our observation sample. The birth masses of the protoplanets are assumed to be $0.01 M_{\oplus}$ and are logarithmically distributed between $0.1$ and $100$\,au. The other distributions of model parameters are summarized in Table~\ref{tab:ca}. For each set of Monte Carlo sampling result, we simultaneously simulate 2000 systems, considering the growth and migration of a single protoplanet.

\begin{table}
    \centering
    \begin{threeparttable}
    \caption{Model parameters in pebble accretion population synthesis study.}
    \label{tab:ca}
    \begin{tabular}{ccl}
        \hline
        Parameter       & Distribution/Value  & 
        Description     \\\hline
        $\dot M_{\rm g0} \ ( M_\odot  \ \rm yr^{-1})$ & $  10^{{\mathcal{N} (\mu, \sigma^2)}} \times (M_{\star}/M_{\odot})^{1.5}$ , $\mu=-6.7$, $\sigma=0.3$ &Disc accretion rate \\
        $R_{\rm d0} \ \rm (AU)$ & $\mathcal{U}$ (50,200) &Disc size \\
        $\rm [Fe/H]$ & $\mathcal{N}$ $(\mu, \sigma^2)$, $\mu=-0.03$, $\sigma=0.2$ &Metallicity \\
        $\xi$ & $0.0149 \times 10^{\mathcal{N} (\mu, \sigma^2)}$, $\mu=-0.03$, $\sigma=0.2$ &Pebble-to-gas flux ratio\\
     $R_{\rm peb}$  &  $1 \ \rm mm$ & Pebbles' radius \\
        \multirow{2}{*}{$\alpha_{\rm g}$} & \multirow{2}{*}{$10^{-2}$} & Global angular momentum \\
& & transport coefficient\\
        $\alpha_{\rm t}$ & $10^{-2}$, $10^{-3}$, $10^{-4}$ & Midplane turbulent strength\\
        $C_1$ & $0.1$, $0.3$, $1.0$ & Migration reduced factor \\
        \hline 
    \end{tabular}
    \begin{tablenotes}
    \item[ ] $\mathcal{N}$($\mu\ ,\ \sigma^{2}$) refers to a normal distribution with mean $\mu$ and standard deviation $\sigma$. 
    \item[ ] $\mathcal{U}$(a\ , \ b) stands for a uniform distribution between $a$ and $b$.
    \end{tablenotes}
    \end{threeparttable}
\end{table}

We summarize the key physical processes and main equations as follows.  The core mass growth proceeds through pebble accretion, with the assumption that pebbles have all grown to a millimeter in size. The corresponding mass growth rate is given by 
 \begin{equation}
 \dot{M}_{\rm PA}  = \varepsilon_{\rm PA} \dot{M}_{\rm peb} = \varepsilon_{\rm PA} \xi_{\rm p/g}  \dot{M}_{\rm g} = \left(\varepsilon_{\rm PA,2D}^{-2} + \varepsilon_{\rm PA,3D}^{-2}\right)^{-1/2}  \xi_{\rm p/g}  \dot{M}_{\rm g}~,
\end{equation}
where $\dot{M}_{\rm peb}$ and $\dot M_{\rm g}$ are the pebble and gas mass flux,  and $\varepsilon_{\rm PA}$ is the total pebble accretion efficiency. The pebble and gas flux ratio is assumed to maintain constant such that $\xi_{\rm p/g}{=}\dot M_{\rm peb} / \dot M_{\rm g}$. The 2D and 3D accretion efficiencies are given by\cite{Liu2018} and\cite{Ormel2018}:
\begin{equation}
        \varepsilon_{\rm PA,2D} = \frac{0.32}{\eta} \sqrt{ \frac{m}{M_{\star}} \frac{1}{\tau_{\rm s}} \frac{\Delta v}{v_{\rm K}}}, \ \  
\varepsilon_{\rm PA,3D} = \frac{0.39}{\eta h_{\rm peb}} \frac{m}{ M_{\star}},
\end{equation}
where $v_{\rm K}{\equiv}\Omega_{\rm K} R$ is the Keplerian velocity, $\Omega_{\rm K}$ is the angular velocity, $R$ is the radial distance to the star,  $\Delta v$ is the relative velocity between the pebbles and planet (dominated by $\eta v_{\rm K}$ in the headwind regime, $\Omega_{\rm K} R_{\rm H}$ in the shear regime, and $R_{\rm H}{=}(m/3M_{\star})^{1/3}R$ is the planet Hill radius),  $h_{\rm peb}$ is the pebble disc aspect ratio, $\eta = -h_{\rm g}^2 (\partial \ln P /\partial \ln R)/2$ and $P$ is the gas disc pressure. 
The pebble disc aspect ratio is $h_{\rm peb}{ =} \sqrt{\alpha_{\rm t}/(\alpha_{\rm t}+\tau_{\rm s}}) \ h_{\rm g}$, and $\tau_{\rm s}{=}\sqrt{2 \pi}\rho_{\bullet} R_{\rm peb}/\Sigma$ is the pebbles' Stokes number, where $\rho_{\bullet}$, $R_{\rm peb}$ and $\Sigma$ are the internal density, radius of pebbles and gas disc surface density. We note that $\Sigma$ and $\dot M_{\rm g}$ are coupled with the global viscous parameter $\alpha_{\rm g}$, whereas $\alpha_{\rm t}$ is the turbulent diffusion coefficient, approximately equivalent to the midplane turbulent strength when the disc is driven by magneto-rotational instability (see detailed two alpha clarification in ref.\cite{Liu2019}). 

We assume that {pebble accretion terminates and gas accretion starts when the planet reaches the pebble isolation mass} $M_{\rm iso}$. The isolation mass formula is based on\cite{Bitsch2018}.  The subsequent gas accretion rate is given by  
\begin{equation}
    \dot{M}_{\rm p, g} = {\rm min} \left[ \left( \frac{{\rm d} M_{\rm p, g}}{{\rm d} t} \right)_{\rm KH}, \ \left( \frac{{\rm d} M_{\rm p, g}}{{\rm d} t} \right)_{\rm Hill}, \ \dot{M}_{\rm g} \right],
\end{equation}
 which is determined by the minimum of the Kelvin-Helmholtz contraction-, the Hill-sphere limited-, and the disc flow limited- accretion rate, respectively. Detailed formulas of each accretion mode can be found in ref.\cite{Liu2019}.

We employ a unified planet migration torque formula from ref.\cite{Kanagawa2018}:
\begin{equation}
    \Gamma  = C_{\rm I} f_{\rm tot} \Gamma_0 = C_{\rm I} \left[ f_{\rm I} f_{\rm s} + f_{\rm II} \left( 1-f_{\rm s} \right) \right] \Gamma_{\rm 0},
\end{equation}
where $\Gamma_{\rm 0} {=} m^2 \, \Sigma \, R^4 \, \Omega_{\rm K}^2/ M_{\star}^2 \, h_{\rm g}^2$ is the normalized torque strength, $f_{\rm I}$ and $f_{\rm II}$ are the type I and type II migration coefficients. The type II migration coefficient $f_{\rm II} {=} -1$ whereas the type I migration coefficient $f_{\rm I}$ is set by the disc thermal structure and local turbulent coefficient $\alpha_{\rm t}$. 
A smooth function of $f_s = 1/[1 + (m/M_{\rm gap})^4]$ is chosen to ensure that $\Gamma{\approx}\Gamma_{\rm I}$ when $m{\ll} M_{\rm gap}$ and $\Gamma{\approx}\Gamma_{\rm I}/(m/M_{\rm gap})^2$ when $m{\gg} M_{\rm gap}$. The gap opening mass $M_{\rm gap}$ is adopted to be $2.3$ times  $M_{\rm iso}$.  
Similar to literature population synthesis studies,  we use a conventional migration reduction factor $C_{\rm I}$ to take into account the uncertainties of this simplified migration prescription.

\begin{figure}
    \centering
    \includegraphics[width=0.8\textwidth]{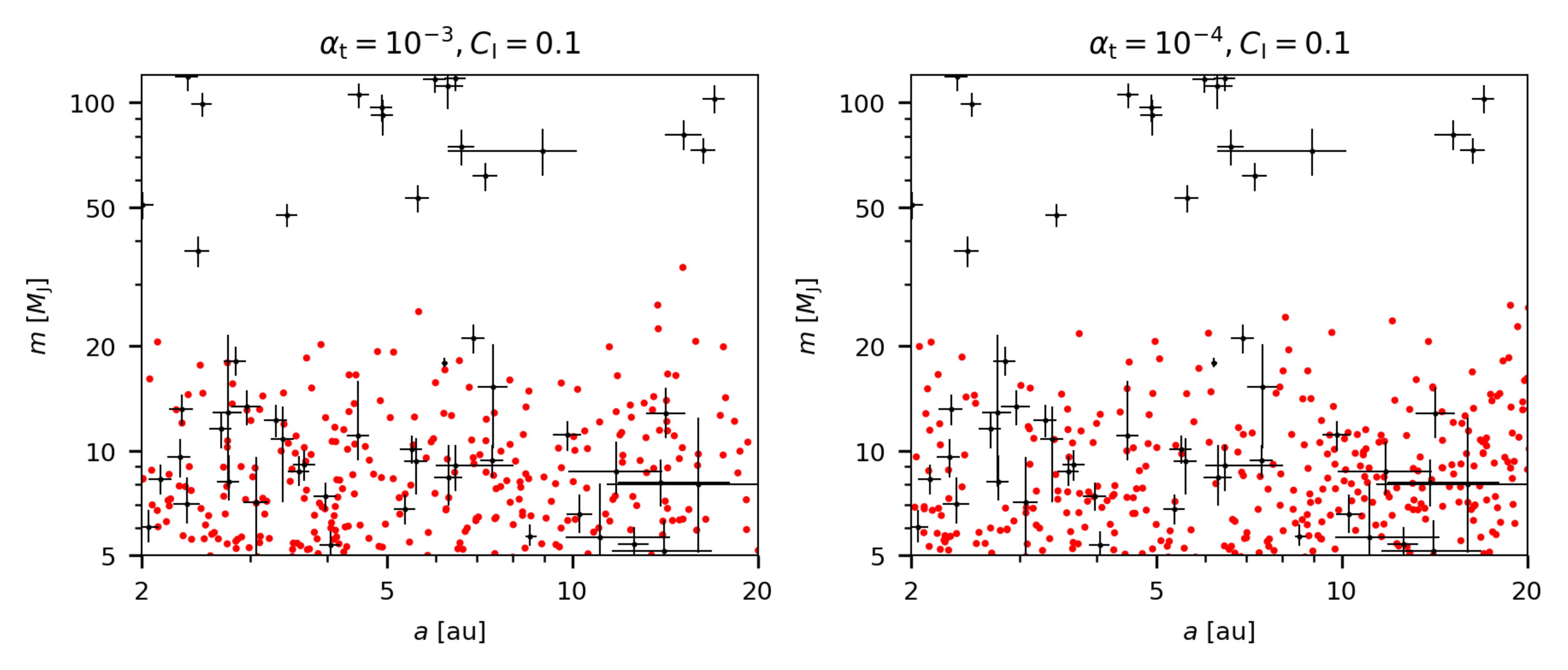}
    \caption{Core accretion populations. The red points are our pebble-driven accretion samples. The black dots are our observation for comparison.}
    \label{fig:ca-pop}
\end{figure}

For detailed model specifications, we refer to ref.\cite{Liu2019}. Particularly, two key model parameters, the mid-plane turbulent viscous strength $\alpha_{\rm t}$ and the migration reduction factor $C_{\rm I}$, are tuned to match the observed exoplanet sample statistically. We explored nine distinct sets of parameters for $\alpha_{\rm t}$ and $C_{\rm I}$. Our findings indicate that when $\alpha_{\rm t} \geq 10^{-3}$ or $\alpha_{\rm t} \geq 10^{-4}$, and $C_{\rm I} = 0.1$, the CA model  successfully accounts for a significant proportion of the observed population of super-Jupiter planets within the snow line.
The Fig. \ref{fig:ca-pop} shows {two simulated population that align the best with observation} with different $\alpha_{\rm t}$.

\subsection*{Gravitational instability model} 
Just as in CA population synthesis, GI population synthesis depends on a number of assumptions\cite{HumphriesEtal19}. 
Here we employ a simplified version of the model focused specifically on how planet migration and gas accretion at different metallicities may shape the resulting population of objects. As in ref.\cite{NayakshinEtal19}, we consider a single planet with an evolving mass $m$ in a  protoplanetary disc that is assumed to conform to $\Sigma  = M_{\rm d}/(\pi R (R_{\rm out} - R_{\rm in}))$ profile, with $R_{\rm in} = 0.1$\,au and $R_{\rm out} = 200$\,au, and $M_{\rm d}$ is the disc mass. We start with $M_{\rm d} = 0.1 M_*$ so that the disc is gravitationally unstable at $t=0$.  The disc mass decreases with time due to three processes: accretion onto the star, disc dispersion modelled with time $t_{\rm disp}$, and accretion of mass onto the planet (fragment). 

Since the emphasis of this paper is on the massive objects only, e.g., very massive gas giant planets and brown dwarfs, here we do not consider possible tidal downsizing of migrating planet embryos\cite{Nayakshin10c, Nayakshin_Review}. For simplicity we consider only one fragment per system in this paper. We inject a fragment with mass $m\sim 1\, \mj$ and let it evolve. The fragment gas accretion is limited by either the Hill capture rate of the gas from the disc, or by the contraction (cooling) of the gas envelope around the planet:
\begin{equation}
    \dot m = \min\left[\frac{m}{t_{\rm acc}}, \dot M_{\rm H}\right]\;,
    \label{dotM_acc_0}
\end{equation}
where $\dot M_{\rm H}$ is given by
\begin{equation}
    \dot M_{\rm H} = \Sigma R_H^2 \Omega_{\rm K},
    \label{dotM_H}
\end{equation}
with Hill radius $R_{\rm H} = R (m/3 M_*)^{1/3}$. 

The planet accretion time scale is physically related to the envelope cooling time, and is introduced as
\begin{equation}
    t_{\rm acc} = t_0 \left( \frac{\mj}{m} \right)^{p_1} \left( \frac{Z}{Z_{\rm sol}} \right)^{p_2}\;,
\end{equation}\label{t_acc_0}
Here $Z$ is the disc metal fraction, assumed to be fixed for each individual simulation, and $Z_{\rm sol}$ is the solar metal fraction, and $t_0$, $p_1$ and $p_2$ are free parameters of the model. Gas accretion onto planets is more efficient at low gas opacity\cite{AyliffeBate09b,AyliffeBate12,Nayakshin17a} since radiative cooling is impeded at high optical depths, and hence we constrain $p_2\geq 0$. By an argument similar to the famous opacity fragmentation limit\cite{Rees76} one can also show that radiative cooling in the envelope of a growing planet is more efficient at higher planet masses, thus $p_1$ is also assumed to be positive.

The approach to planet migration is similar to that in ref.\cite{Nayakshin20-Paradox}. We set the disc temperature profile to $T = T_0 (R_0/R)^{1/2}$, where $T_0$ = 20 K, in line with values typically used in disc models and ALMA discs\cite{LongEtal18}. The geometric aspect ratio of the disc model is therefore
\begin{equation}
    \frac{H}{R} = \left( \frac{k_{\rm B} T R}{G M_* \mu} \right)^{1/2} = 0.11 \left( \frac{R}{R_0} \right)^{1/4}  \left( \frac{0.7 \mathrm M_{\odot}}{M_*}\right)^{1/2},
    \label{h_r}
\end{equation}
where $H$ is the vertical scale height of the disc, $k_{\rm B}$ is Boltzmann's constant and $\mu = 2.45\,m_{\rm p}$ is the mean molecular weight, where $m_{\rm p}$ is the proton mass. Planets undergo either the type I (no deep gap is opened in the disc) or the type II (deep gap is opened within the disc) migration, following the well known Crida\cite{CridaEtal06} criterion. A deep gap is opened in the disc, and planet migration switches from type I to type II, when the Crida parameter
\begin{equation}
    C = \frac{3 H}{4 R_{\rm H}} + 50\,\alpha_{\rm v} \left( \frac{H}{R}\right)^2 \frac{M_*}{m} < 1,
    \label{cp}
\end{equation}
were $ \alpha_{\rm v} $ is the\cite{Shakura73} disc viscosity parameter.

The type I migration time for companion of mass $m$ is given by
\begin{equation}
    t_{\rm mig 1} = \frac{1}{2 \gamma \Omega_{\rm K}} \frac{M_*^2}{m \Sigma R^2} \left( \frac{H}{R} \right)^2,
    \label{tmig1_1}
\end{equation}
where 
$\gamma$ is dimensionless and depends on disc properties\cite{PaardekooperEtal10a}, evaluating to \( \gamma = 2.5\) in our disc model. Equation \ref{tmig1_1} can also be expressed as
\begin{equation}
    t_{\rm mig 1} = 7 \times 10^4 \mathrm{yr} \left( \frac{R}{R_0} \right)^{2} \left( \frac{M_*}{0.7 \mathrm M_{\odot}} \right)^{1/2} \left( \frac{1 \mathrm M_{\rm J}}{m} \right)\frac{10 \mathrm M_{\rm J}}{M_{\rm loc}}.
    \label{tmig1}
\end{equation}
 When $C<1$, the type II planet migration applies, which is described by
\begin{equation}
    t_{\rm mig2} = \frac{1}{\alpha_{\rm v} \Omega} \left( \frac{R}{H}\right)^2 \left( 1 + \frac{m}{4 \pi \Sigma R^2}\right)
\end{equation}

\subsection*{combined population synthesis}
In our CA model, only two sets of parameters ($\alpha_{\rm t} = 10^{-3}$, $C_{\rm I} = 0.1$ and $\alpha_{\rm t} = 10^{-4}$, $C_{\rm I} = 0.1$) are valid as discussed above. Consequently, we select these as our candidate CA models. For each candidate CA model, we subsequently adjust the parameters within the GI model to achieve an optimal fit.

In our GI model, as described in the above content, we have six free parameters represented by $\boldsymbol{\theta}_\mathrm{GI}(\alpha, M_{\rm d}^{\prime}, M_{\rm p}^{\prime}, t_\mathrm{disp}, p_1, p_2)$. The viscosity parameter $\alpha$, the power index of planetary mass $p_1$, and the power index of the metallicity $p_2$ are constants for all stellar sample. The terms $M_d$, $M_p$, and $t_\mathrm{disp}$ serve as hyper-parameters governing the initial distributions in our simulations.
The initial disc mass distribution follows a truncated normal distribution with the mean value $\mu=M_d$, standard deviation $\sigma=0.1$, the lower and upper bounds are set to 0.05 and 1 in the unit of $M_\odot$. The initial planetary mass follows a uniform distribution range from $M_p$ to $4M_p$ in the unit of $\mj$. The disc dispersion time scale in the unit of year, when expressed as the base-10 logarithm follows a truncated normal distribution with the $\mu=\lg t_\mathrm{disp}$, $\sigma=\lg t_\mathrm{disp}$, the lower and upper bounds are set to 4.5 and 6.5. 

We also have some fixed parameters in GI model. The total integration time is defined as $t_\mathrm{max} = 10^5 + 3 t_\mathrm{disp}$. We characterize the initial disc radius using a logarithmic scale, where $\lg R_\mathrm{disc}$ adheres to a truncated normal distribution with a mean of 2 and a standard deviation of 0.5. This distribution is constrained within the range of [1,3], which translates to physical disc radii spanning from 10 au to 1000 au. Additionally, we adopt a self-gravity induced viscosity parameter $\alpha_{sg}$ with a value of 0.04.

\begin{figure}
    \centering
    \includegraphics[width=0.5\linewidth]{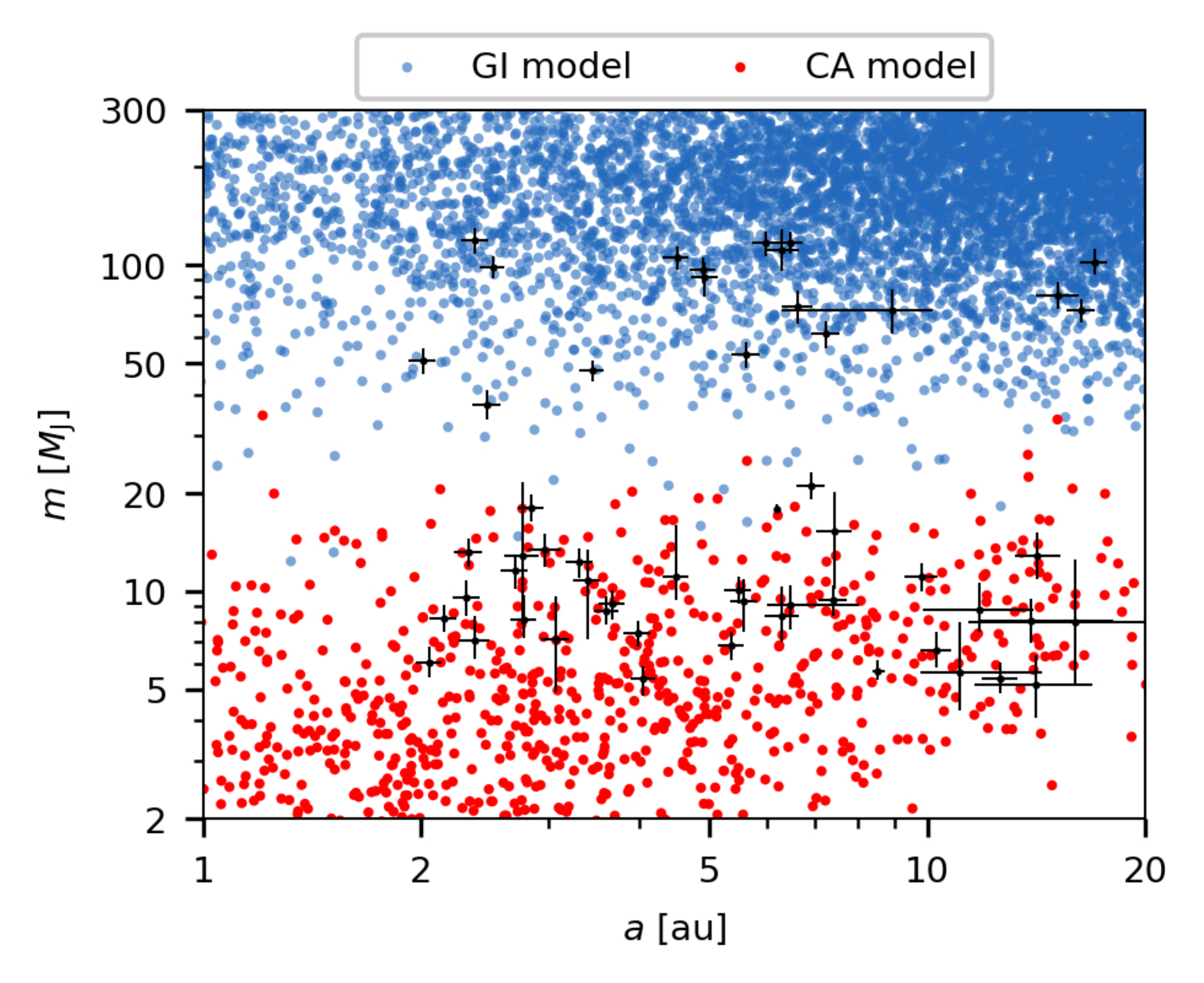}
    \caption{Result of the combined population synthesis. Populations derived from the GI model are shown in {blue}, while those resulting from the CA model are depicted in {red}. Observational data are represented by black points.}
    \label{fig:pop-syn}
\end{figure}

\begin{figure}
    \centering
    \includegraphics[width=0.5\linewidth]{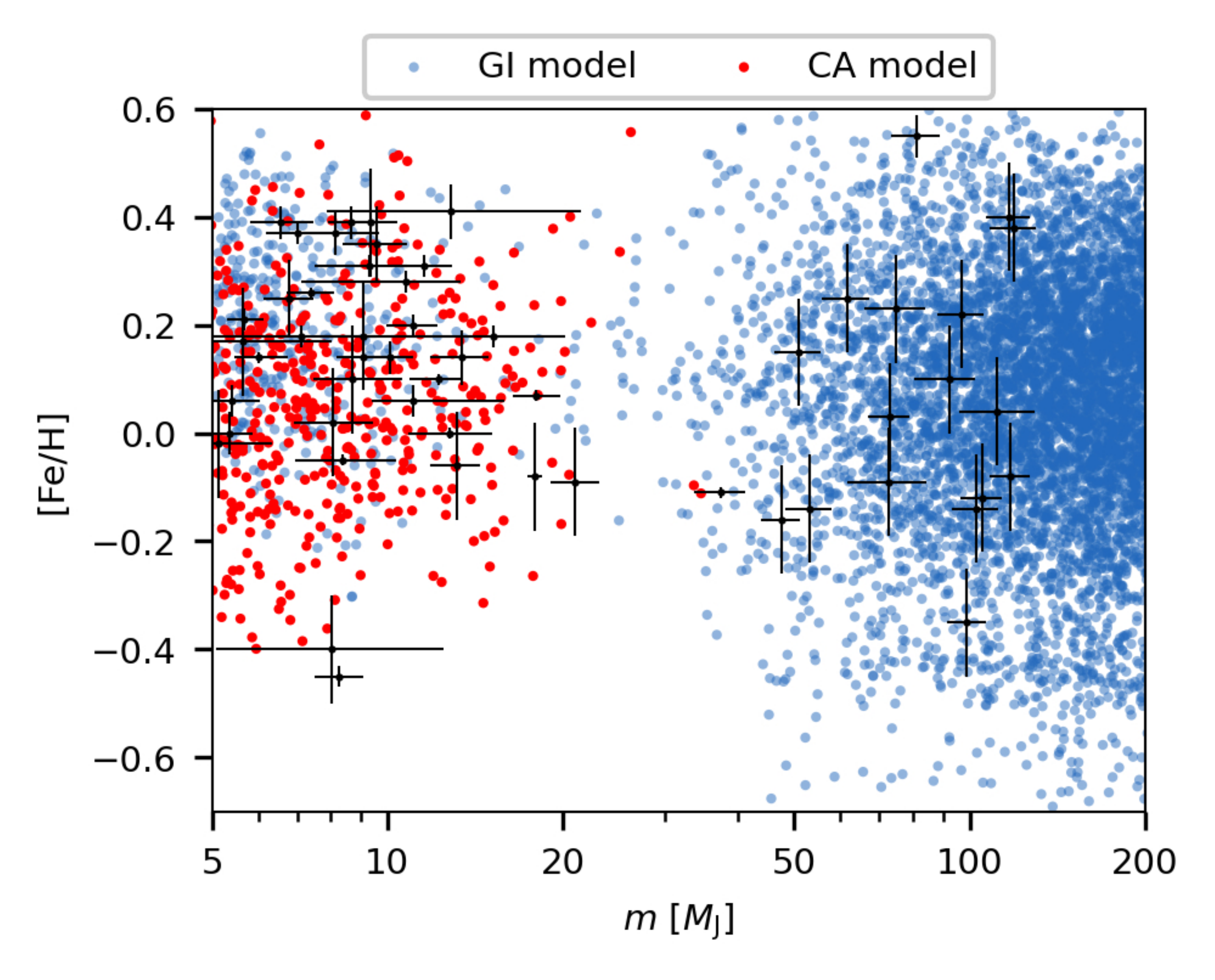}
    \caption{Mass and metallicity distributions of the combined population synthesis. Markers and colors are consistent with those in Extended Fig. \ref{fig:pop-syn}.}
    \label{fig:mass-meta-dist}
\end{figure}

After fitting with observation, the CA model with $\alpha_{\rm t} \geq 10^{-3}$ qualitatively aligns most closely with observations. The parameters of combined model with their one-sigma errors are presented in Table~\ref{tab:best-pop-model-parameters}. Fig.~\ref{fig:pop-syn} illustrates the distinct distributions of companions formed via CA and GI within the $a\text{--}m$ plane. It is also clear that GI-formed companions are preferentially found at lower metallicities, particularly among higher-mass companions (shown in the Fig.~\ref{fig:mass-meta-dist}).

\begin{table}
\centering
\caption{Best-fit parameters of the combined population synthesis model}
\begin{tabular}{ccccccccc} 
\hline
Parameter & $\lg\alpha$     & $M_d$ & $M_p$ & $f_1$ & $f_2$ & $\lg t_\mathrm{disp}$ & $p_1$ & $p_2$  \\ 
\hline
Value     & $-1.2\pm0.2$ & $0.2\pm0.1$  & $4.5\pm0.4$  & $0.08\pm0.001$ & $0.04\pm0.001$ & $6.4\pm0.1$  & $0.9\pm0.3$  & $3.0\pm0.2$   \\
\hline
\end{tabular}
\label{tab:best-pop-model-parameters}
\end{table}

\section*{Software}\label{sec:software}
We acknowledge the teams who developed following open-source software packages used in this work: \texttt{astropy} \citep{theastropycollaborationAstropyCommunityPython2013,theastropycollaborationAstropyProjectBuilding2018}, 
\texttt{numpy} \citep{harrisArrayProgrammingNumPy2020}, \texttt{pandas} \citep{mckinney-proc-scipy-2010,reback2020pandas}, 
\texttt{scipy} \citep{2020SciPy-NMeth}, 
\texttt{matplotlib} \citep{hunterMatplotlib2DGraphics2007},
\texttt{emcee} \citep{foreman-mackeyEmceeMCMCHammer2013}.

\end{document}